\documentclass[aps,prx,twocolumn,superscriptaddress,amsmath,amssymb,nofootinbib]{revtex4-2}
\pdfoutput=1
\usepackage{graphicx}  
\usepackage{amsmath}
\usepackage[dvipsnames]{xcolor}
\usepackage{dcolumn}   
\usepackage{bm}        
\usepackage{verbatim}   
\usepackage{gensymb}
\usepackage{epstopdf}
\usepackage{footmisc}
\usepackage{natbib}

\usepackage{color}
\usepackage{subcaption} 
\usepackage{multirow}
\usepackage{physics}
\usepackage{braket}
\usepackage{mathtools}
\usepackage{threeparttable}
\usepackage[referable]{threeparttablex}
\usepackage{hyperref}

\newcommand{\fwtf}{$m_{\mathrm{wtf}}$}
\newcommand{\fl}{$m_{\mathrm{L}}$}
\newcommand{\MST}{MnSb$_2$Te$_4$}
\newcommand{\MBT}{MnBi$_2$Te$_4$}
\newcommand{\MBTw}{MnBi$_4$Te$_7$}
\newcommand{\MBTh}{MnBi$_6$Te$_{10}$}
\newcommand{\MBTnn}{(MnBi$_2$Te$_4$)(Bi$_2$Te$_3$)$_n$}
\newcommand{\MBTnl}{Mn$_2$Bi$_2$Te$_5$}
\newcommand{\MBS}{MnBi$_2$Se$_4$}
\newcommand{\Mn}{$^{55}$Mn}
\newcommand{\Mnc}{Mn$_{\mathrm 6c}$}
\newcommand{\Mna}{Mn$_{\mathrm 3a}$}
\newcommand{\muSR}{$\mu$SR}
\newcommand{\muB}{\mu_B}
\newcommand{\Bia}{Bi$_{\mathrm 3a}$}

\newcommand{\BiTe}{Bi$_2$Te$_3$}
\newcommand{\TC}{$T_{\mathrm{C}}$}
\newcommand{\TN}{$T_{\mathrm{N}}$}
\newcommand{\Tm}{$T_{\mathrm{m}}$}
\newcommand{\Tstar}{$T^*$}
\newcommand{\Bhf}{$B_{\mathrm{hf}}$}
\newcommand{\BMn}{$^{55}B$}
\newcommand{\Bmu}{$B_\mu$}
\newcommand{\BL}{$B_{\mathrm{L}}$}
\newcommand{\B}{$\mu_0 H$}
\newcommand{\DFTm}{DFT+$\mu$}
\newcommand{\TM}{Te-$\mu$-Mn}
\newcommand{\TB}{Te-$\mu$-Bi}
\newcommand{\TMB}{Te-$\mu$-Bi@Mn}
\newcommand{\TBM}{Te-$\mu$-Mn@Bi}

\newcommand{\TT}{Te-$\mu$-Te}
\newcommand{\pB}{$p(B_\mu)$}

\begin{document}

\title{Ubiquitous order-disorder transition in the Mn antisite sublattice of the \MBTnn\ magnetic topological insulators}

\author{M.~Sahoo}
\thanks{MS and IJO contributed equally}
\affiliation{Leibniz IFW Dresden, Helmholtzstra\ss{}e 20, D-01069 Dresden, Germany}
\affiliation{Institut f\"{u}r Festk\"{o}rper- und Materialphysik, Technische Universit\"{a}t Dresden, 01062 Dresden, Germany}
\affiliation{Würzburg-Dresden Cluster of Excellence ct.qmat, Germany}
\affiliation{Dipartimento di Scienze Matematiche, Fisiche e Informatiche, Universit\'a di Parma, Parco delle Scienze 7A, I-43124 Parma, Italy}

\author{I.J.~Onuorah}
\thanks{MS and IJO  contributed equally}
\affiliation{Dipartimento di Scienze Matematiche, Fisiche e Informatiche, Universit\'a di Parma, Parco delle Scienze 7A, I-43124 Parma, Italy}

\author{L.C.~Folkers}
\affiliation{Institut f\"{u}r Festk\"{o}rper- und Materialphysik, Technische Universit\"{a}t Dresden, 01062 Dresden, Germany}
\affiliation{Würzburg-Dresden Cluster of Excellence ct.qmat, Germany}


\author{E.V.~Chulkov}
\affiliation{Donostia International Physics Center, 20018 Donostia-San Sebastián, Spain}
\affiliation{Departamento de Polímeros y Materiales Avanzados: Física, Química y Tecnología, Facultad de Ciencias Químicas, Universidad del País Vasco UPV/EHU, 20018 Donostia-San Sebastián, Spain}
\affiliation{Centro de Física de Materiales (CFM-MPC), Centro Mixto (CSIC-UPV/EHU), 20018 Donostia-San Sebastián, Spain}
\affiliation{Saint Petersburg State University, 199034 Saint Petersburg, Russia}
\author{M.M. Otrokov}
\affiliation{Instituto de Nanociencia y Materiales de Aragón (INMA), CSIC-Universidad de Zaragoza, 50009 Zaragoza, Spain}
\author{Z.S. Aliev}
\affiliation{Baku State University, AZ1148 Baku, Azerbaijan}
\affiliation{Institute of Physics Ministry of Science and Education Republic of Azerbaijan, AZ1143 Baku, Azerbaijan}
\author{I.R.~Amiraslanov}
\affiliation{Baku State University, AZ1148 Baku, Azerbaijan}
\affiliation{Institute of Physics Ministry of Science and Education Republic of Azerbaijan, AZ1143 Baku, Azerbaijan}
\author{A.U.B.~Wolter}
\affiliation{Leibniz IFW Dresden, Helmholtzstra\ss{}e 20, D-01069 Dresden, Germany}

\author{B.~Büchner}
\affiliation{Leibniz IFW Dresden, Helmholtzstra\ss{}e 20, D-01069 Dresden, Germany}
\affiliation{Institut f\"{u}r Festk\"{o}rper- und Materialphysik, Technische Universit\"{a}t Dresden, 01062 Dresden, Germany}
\affiliation{Würzburg-Dresden Cluster of Excellence ct.qmat, Germany}

\author{L.~T.~Corredor}
\affiliation{Leibniz IFW Dresden, Helmholtzstra\ss{}e 20, D-01069 Dresden, Germany}

\author{Ch.~Wang}
\address{Laboratory for Muon Spin Spectroscopy, Paul-Scherrer-Institute, CH-5232 Villigen PSI, Switzerland.
}

\author{Z.~Salman}
\address{Laboratory for Muon Spin Spectroscopy, Paul-Scherrer-Institute, CH-5232 Villigen PSI, Switzerland.
}

\author{A.~Isaeva}
\affiliation{Leibniz IFW Dresden, Helmholtzstra\ss{}e 20, D-01069 Dresden, Germany}
\affiliation{Van der Waals-Zeeman Institute, Department of Physics and Astronomy, University of Amsterdam, Science Park 094, 1098 XH Amsterdam, The Netherlands}

\author{R. De Renzi}
\email[Corresponding address: ]{roberto.derenzi@unipr.it}
\affiliation{Dipartimento di Scienze Matematiche, Fisiche e Informatiche, Universit\'a di Parma, Parco delle Scienze 7A, I-43124 Parma, Italy}
\author{G.~Allodi}
\affiliation{Dipartimento di Scienze Matematiche, Fisiche e Informatiche, Universit\'a di Parma, Parco delle Scienze 7A, I-43124 Parma, Italy}

\date{\today}
\begin{abstract}
Magnetic topological insulators (TIs) herald a wealth of applications in spin-based technologies, relying on the novel quantum phenomena provided by their topological properties. Particularly promising is the (MnBi$_2$Te$_4$)(Bi$_2$Te$_3$)$_n$ layered family of established intrinsic magnetic TIs that can flexibly realize various magnetic orders and topological states. High tunability of this material platform is enabled by manganese--pnictogen intermixing, whose amounts and distribution patterns are controlled by synthetic conditions. 

Positive implication of the strong intermixing in  MnSb$_2$Te$_4$ is the interlayer exchange coupling switching from antiferromagnetic to ferromagnetic, and the increasing magnetic critical temperature. On the other side, intermixing also implies atomic disorder which may be detrimental for applications. Here, we employ nuclear magnetic resonance and muon spin spectroscopy, sensitive local probe techniques, to scrutinize the impact of the intermixing on the magnetic properties of \MBTnn~and \MST. Our measurements not only confirm the opposite alignment between the  Mn magnetic moments on native sites and antisites in the ground state of \MST, 
but for the first time directly show the same alignment in \MBTnn\ with $n=0, 1$ and 2. Moreover, for all compounds, we find the static magnetic moment of the Mn antisite sublattice to disappear well below the intrinsic magnetic transition temperature, leaving a homogeneous magnetic structure undisturbed by the intermixing.
Our findings provide a microscopic understanding of the crucial role played by Mn--Bi intermixing in \MBTnn \ and offer pathways to optimizing the magnetic gap in its surface states.

\end{abstract}
\maketitle

\date{\today}

\maketitle
\section{Introduction}
The interplay between non-trivial topology and magnetic order has been under the spotlight since the advent of a topological era in condensed matter physics because it may enable versatile and tunable topological phases~\cite{Qi2011, tokura2019magnetic, wang2021intrinsic, Bernevig2022, Chang.rmp2023}. Magnetic topological 
materials emerged as an ideal platform for harbouring emergent 
quantum phenomena of technological relevance, including the quantum anomalous Hall effect (QAHE) and axion electrodynamics~\cite{chang2013experimental, Checkelsky.nphys2014, Nenno2020axion, Sekine2021axion, Qiu.nmat2023}. A layered (van der Waals) compound  \MBT, which is a magnetic derivative 
of the prototypical \BiTe~topological insulator, has been established as the first intrinsically magnetic~TI~\cite{otrokov2019prediction, li2019intrinsic, Zhang.prl2019, Gong.cpl2019}. In this compound, the local moments of the Mn atoms adopt
an A-type antiferromagnetic (AFM) order, consisting of a ferromagnetic (FM) alignment within the Mn layer, with AFM stacking in the perpendicular 
direction \cite{otrokov2019prediction, Gong.cpl2019, Yan.prm2019}. 
Combination of a layered crystal structure and the A-type AFM order makes \MBT\, to fall under the $Z_2$ topological classification of AFM insulators \cite{Mong.prb2010}. The non-trivial value of the invariant, $Z_2=1$, stemming from the spin-orbit coupling driven bulk band gap inversion, categorizes it as a three-dimensional AFM TI~\cite{otrokov2019prediction, li2019intrinsic, Zhang.prl2019, Gong.cpl2019}. In the two-dimensional limit, thin \MBT\, films were theoretically predicted~\cite{Otrokov.prl2019, Li.sciadv2019} and experimentally confirmed to show the AFM axion insulator state~\cite{liu2020robust, Gao.nat2021} as well as the quantized Hall effect, both under external field~\cite{liu2020robust, Gao.nat2021, Ge.nsr2019, deng2020quantum} and in remanence~\cite{deng2020quantum}.

 On top of its exciting intrinsic properties, \MBT\, also fosters a highly tunable material platform. Multiple tuning knobs, not only extrinsic such as magnetic field, pressure, and temperature, but also intrinsic such as Mn--Mn interlayer distance, variations of the chemical composition, and defect engineering, result in various magnetic and topological states. For example, various pnictogen or chalcogen substitutions and Mn/Bi/Te stoichiometry alternations give rise to  such materials as \MST, \MBS, or \MBTnl, whose magnetic and topological properties have been studied both theoretically and experimentally \cite{Yan.prb2019, wimmer2021mn, Lee.prx2021, Zhu.nl2021, Cao.prb2021}. Furthermore, the van der Waals nature of \MBT\, tolerates interlacing the adjacent  (\MBT) septuple layers with various number $n$ of nonmagnetic (\BiTe) quintuple layers, resulting in the  \MBTnn~family of stacked structures ($n=1$ for \MBTw,  $n=2$ for \MBTh, etc.)~\cite{Aliev.jac2019, Souchay.jmcc2019}. The increasing distance between the septuple layers progressively weakens the interlayer exchange coupling with an increasing $n$, which enables an effective tuning of the magnetic structure by moderate magnetic fields~\cite{vidal2019topological, wu2019natural, hu2020van, klimovskikh2020tunable}, or hydrostatic pressure~\cite{shao2021pressure}, driving these compounds from the AFM to the FM state. The \MBTnn\ materials may host exotic, field-induced topological phases~\cite{Zhang.prl2020, Hu.sciadv2020, vidal2019topological} and temperature-dependent metamagnetic states~~\cite{tan2020metamagnetism}.
 
Native defects in these materials are lately in the center of attention thanks to their strong influence on the magnetic and electronic structure~\cite{PhysRevX.11.021033, Lai2021, Garnica.npjqm2022, Liu.pnas2022, Tan.prl2023, Lupke.comsmat2023}. They are exploited as an effective tuning knob to purposely modify the latter~\cite{PhysRevX.11.021033, tcakaev2023intermixing, Yan.nl2022, SAHOO2023101265}. These defects originate from antisite intermixing between the native manganese and pnictogen crystallographic sites. This phenomenon is favored by similar ionic radii, especially those of the Mn and Sb~\cite{folkers2022occupancy}, which enables ca.~3 times stronger degree of intermixing in \MST\ compared to \MBT. Specifically, Mn atoms partially occupy the Sb/Bi $6c$ Wyckoff  site (\Mnc\ antisite), while pnictogen atoms swap to the Mn $3a$ site (\Bia\ antisite). The amounts of swapped cations do not necessarily fulfil the electroneutrality assumptions for Mn(II) and Sb(III)/Bi(III), and the occurrence of cationic vacancies in both sites is debated~\cite{defects}. Strong intermixing in \MST\, promotes the FM interlayer coupling~\cite{PhysRevB.100.195103}. The magnetic transition temperature jumps from \TN=19 K in the AFM-like bulk Mn$_{1-x}$Sb$_{2+x}$Te$_4$ single crystals~\cite{PhysRevX.11.021033} to \TC=58 K in the FM-like Mn$_{1+x}$Sb$_{2-x}$Te$_4$ ones~\cite{SAHOO2023101265}, which is achieved by varying the growth conditions. (Hereinafter, the main magnetic transition is referred to as \Tm, meaning either \TN\ for AFM or \TC\ for FM samples.) Moreover, the interlayer coupling can become truly FM in \MBTh\ via Mn-Bi defects engineering under appropriate growth conditions~\cite{tcakaev2023intermixing, Yan.nl2022}. This FM coupling may not only help to realize an FM axion insulator state~\cite{Zhang.prl2020, Hu.sciadv2020}, but also the QAHE in contrast to the AFM \MBT\ \cite{deng2020quantum, Chong.arxiv2023}.

Low-temperature neutron diffraction measurements, performed on both the AFM- and FM-like \MST\ bulk single crystals, reveal that the local moments of the Mn$_{6c}$ atoms are coupled antiparallel to the Mn$_{3a}$ ones (\emph{ferri}magnetic structure)~\cite{PhysRevX.11.021033, Riberolles2021}. However, for the \MBTnn\ family the magnetic role of the antisites has not been decisively established yet. Indeed, the available neutron diffraction data \cite{Yan.prm2020, Ding.prb2020-2, Ding.jpd2021} do not shed light on this issue (likely because of the much lower levels of intermixing as compared to \MST), although the high-field magnetization studies performed on \MBT\ do suggest that there is an AFM coupling between the Mn$_{6c}$ and Mn$_{3a}$ sublattices \cite{Lai2021}.

To close this important gap, this article systematically investigates the magnetic behavior of the antisites in polycrystalline  \MBTnn\ with $n=0,1,2$ and in one \MST\, sample  by means of local magnetic probes, namely nuclear magnetic resonance (NMR) (Sec.~\ref{sec:NMR}) and muon spin spectroscopy ($\mu$SR) (Sec.~\ref{sec:muSR}). Both techniques probe the dynamic and thermodynamic material properties, as well as the disorder introduced by the antisites. The measurements are performed in applied and in zero (ZF) external magnetic field, and \Mn~NMR provides direct local evidence of the nearly opposite relative spin alignment of \Mna~and \Mnc~for all compounds studied here. 
Importantly, NMR reveals a magnetic order-disorder transition in the \Mnc~sublattice of \MST\  well below $T_{\mathrm{m}} = 27$~K. 
This clearly discriminates two regions 
in the \MST\ magnetic phase diagram: (i) $T<T^*<$ \Tm, when the \Mnc\ and \Mna\  moments are coupled antiferromagnetically; and (ii) $T^*<T<$ \Tm, when the \Mnc\ sublattice is paramagnetic-like, whereas the \Mna\ one sustains its intra- and interlayer 
orders. \muSR~and bulk magnetometry measurements confirm the same order-disorder transition in the \Mnc~sublattice of all studied \MBTnn~as well. 

The discovery of the loss of the \Mnc\ sublattice magnetic ordering is highly relevant in the context of the crucial role that Mn-Bi intermixing plays in the reduction of the \MBT\ Dirac point gap \cite{Garnica.npjqm2022, Liu.pnas2022, Tan.prl2023, Lupke.comsmat2023, 
Hao.prx2019, Li.prx2019, Chen.prx2019, Swatek.prb2020, Nevola.prl2020, Shikin.prb2021}.

\section{The antisites from the NMR perspective}
\label{sec:NMR}

In this and the next section, we describe a unified picture emerging from NMR and \muSR, skipping the technical details of its derivation for the sake of clarity. The principles of the two techniques are briefly described in Sec.~\ref{sec:methods} and further important details are provided in the Supplemental Material \cite{SM}. Sec.~\ref{sec:methods} also gives details about the sample preparation and characterization and is based on our previous works~\cite{Zeugner.cm2019, vidal2019topological, tcakaev2023intermixing, folkers2022occupancy}.

In ZF-NMR, \Mn\ nuclear spins (gyromagnetic ratio ${^{55}\gamma}=10.576$ MHz/T) precess around a very large hyperfine field 
\Bhf, 
between 40 and 45 T at 1.4 K. \Bhf~is primarily due 
to the negative coupling, $ -\cal A$, to the on-site moment $g\muB\mathbf{S}$, with much smaller positive couplings $\cal B$ to six nearest neighbors \Mna, and even smaller distant dipole contributions, so that in first approximation, assuming parallel $\mathbf{S}$ in the layer, the NMR frequency is ${^{55}\nu} = {^{55}\gamma}({\cal A}-6{\cal B}) g \mu_{\mathrm{B}}\mathbf{S}$. 

\begin{figure*}
\centering
\begin{subfigure}{.25\textwidth}
  \includegraphics[width=0.95\linewidth]{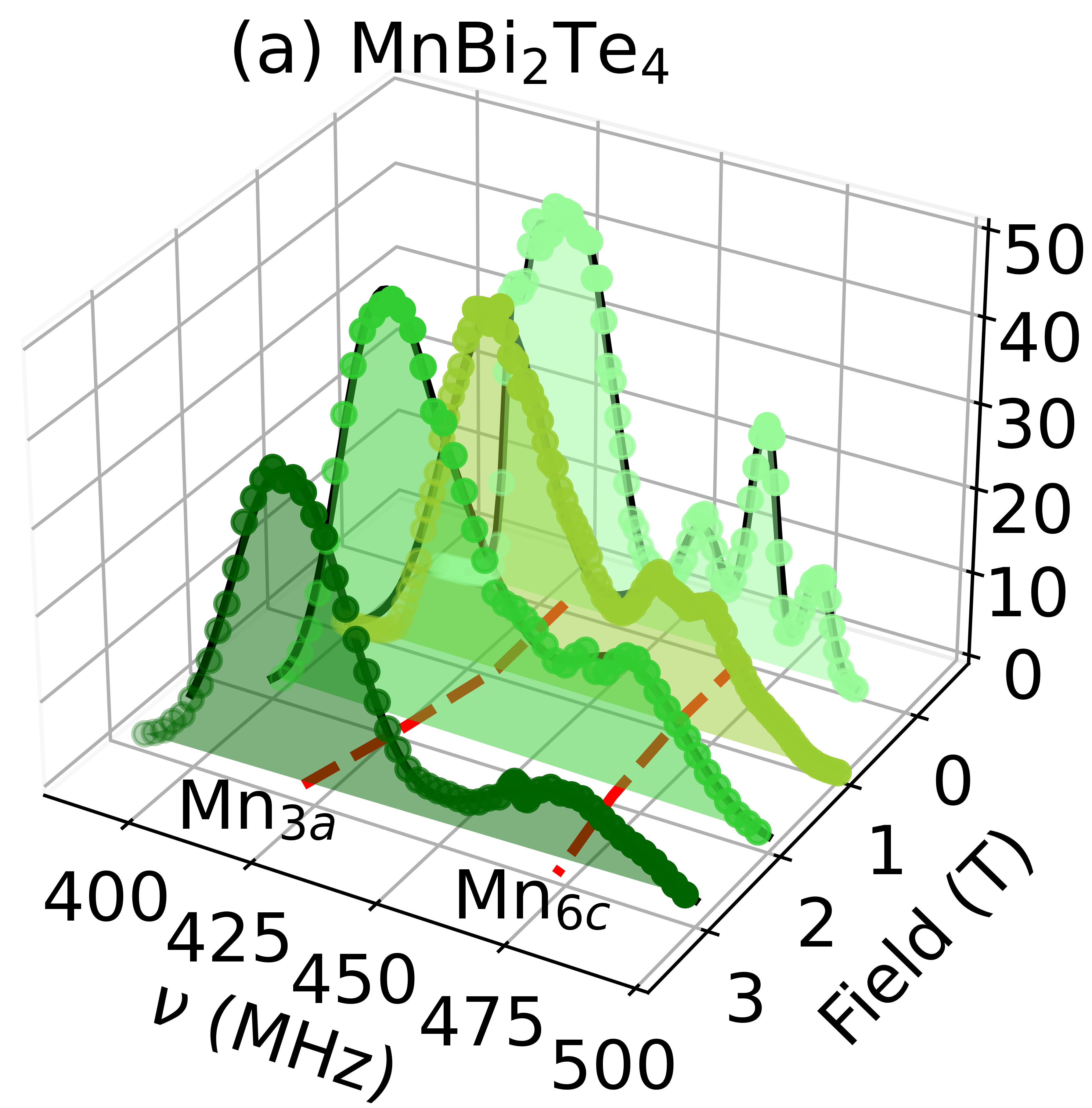}
  \label{fig:sfig1}
\end{subfigure}%
\begin{subfigure}{.25\textwidth}
  \includegraphics[width=0.95\linewidth]{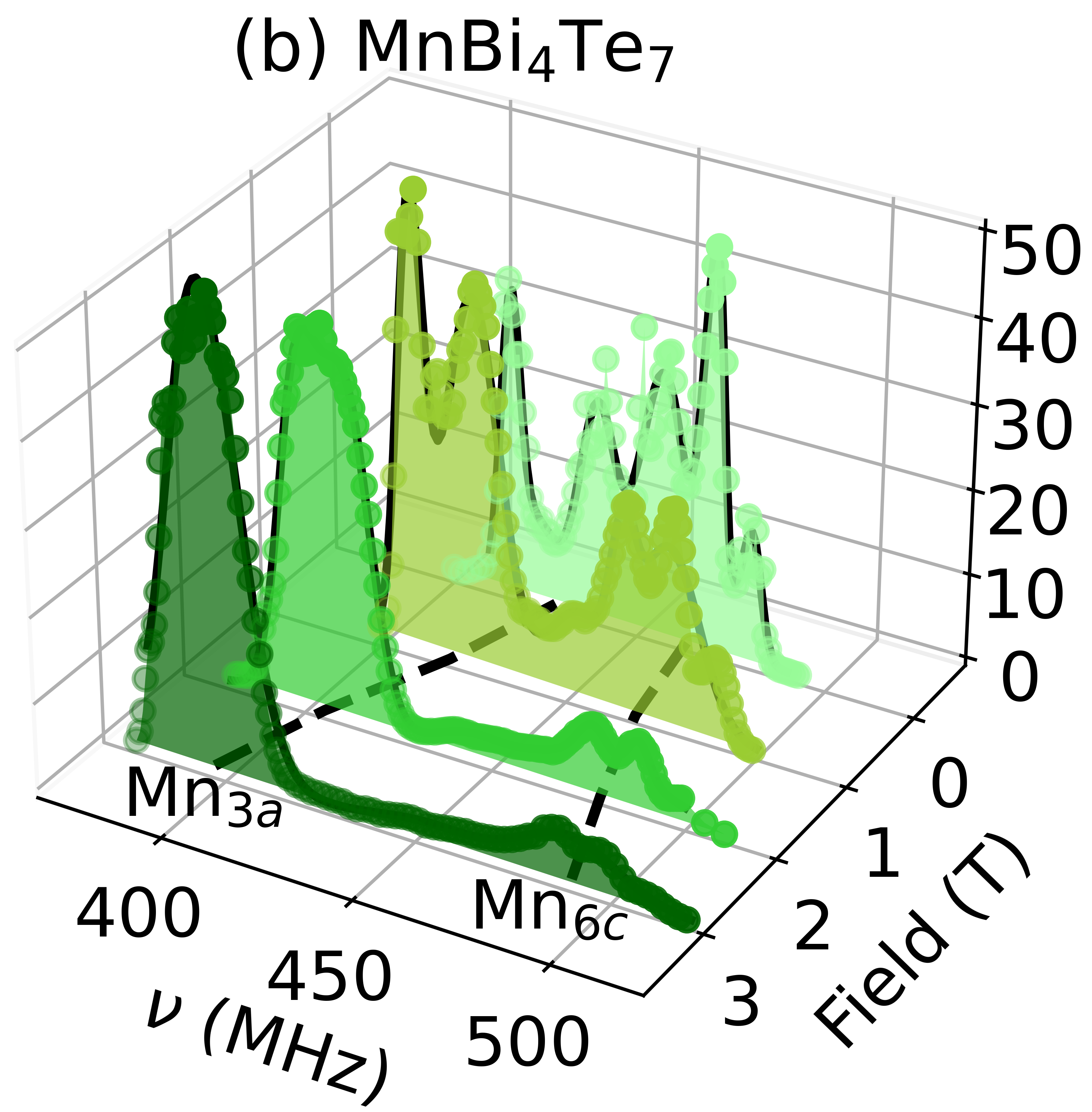}
  \label{fig:sfig2}
\end{subfigure}
\begin{subfigure}{.25\textwidth}
  \includegraphics[width=0.95\linewidth]{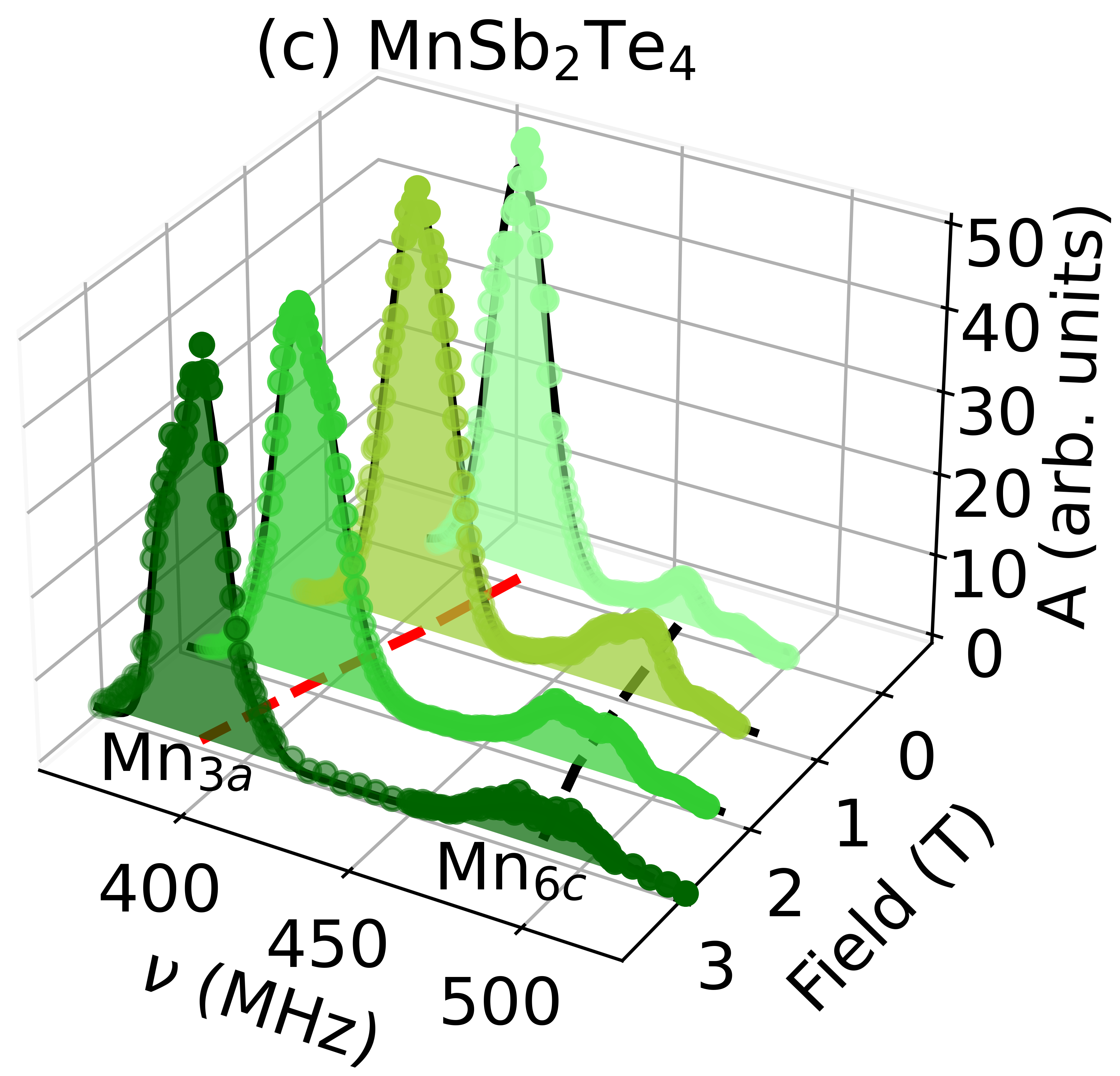}
  \label{fig:sfig1t}
\end{subfigure}
\begin{subfigure}{.22\textwidth}
  \centering
  \includegraphics[width=0.95\linewidth]{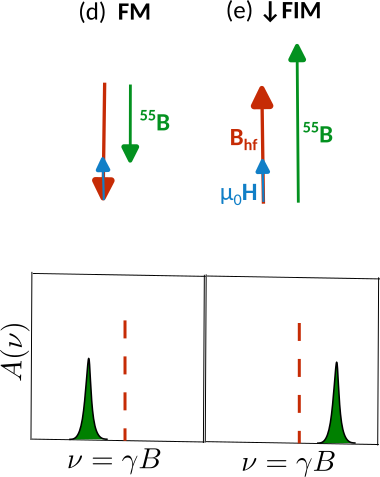}
  \label{sfig:fmfim}
\end{subfigure}%

\begin{subfigure}{.23\textwidth}
  \centering
  \includegraphics[width=0.95\linewidth]{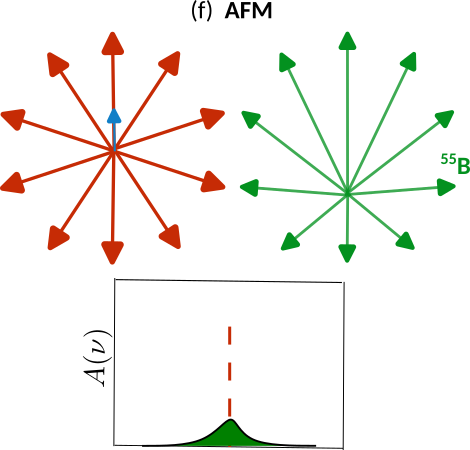}
  \label{sfig:afm}
\end{subfigure}%
\begin{subfigure}{.35\textwidth}
  \centering
  \includegraphics[width=1.0\linewidth]{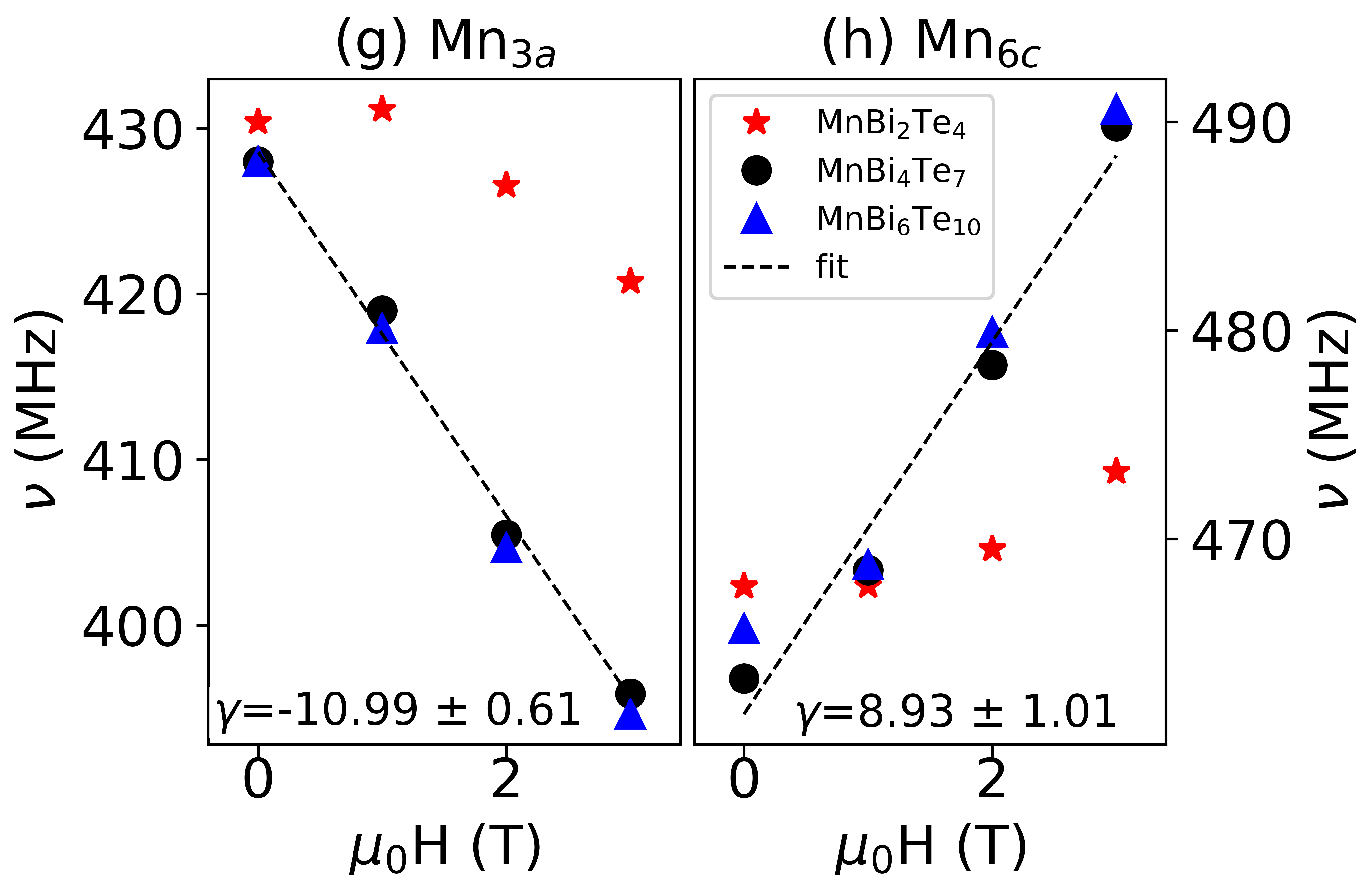}
  \label{fig:sfig4}
\end{subfigure}%
\begin{subfigure}{.35\textwidth}
  \centering
  \includegraphics[width=1.0\linewidth]{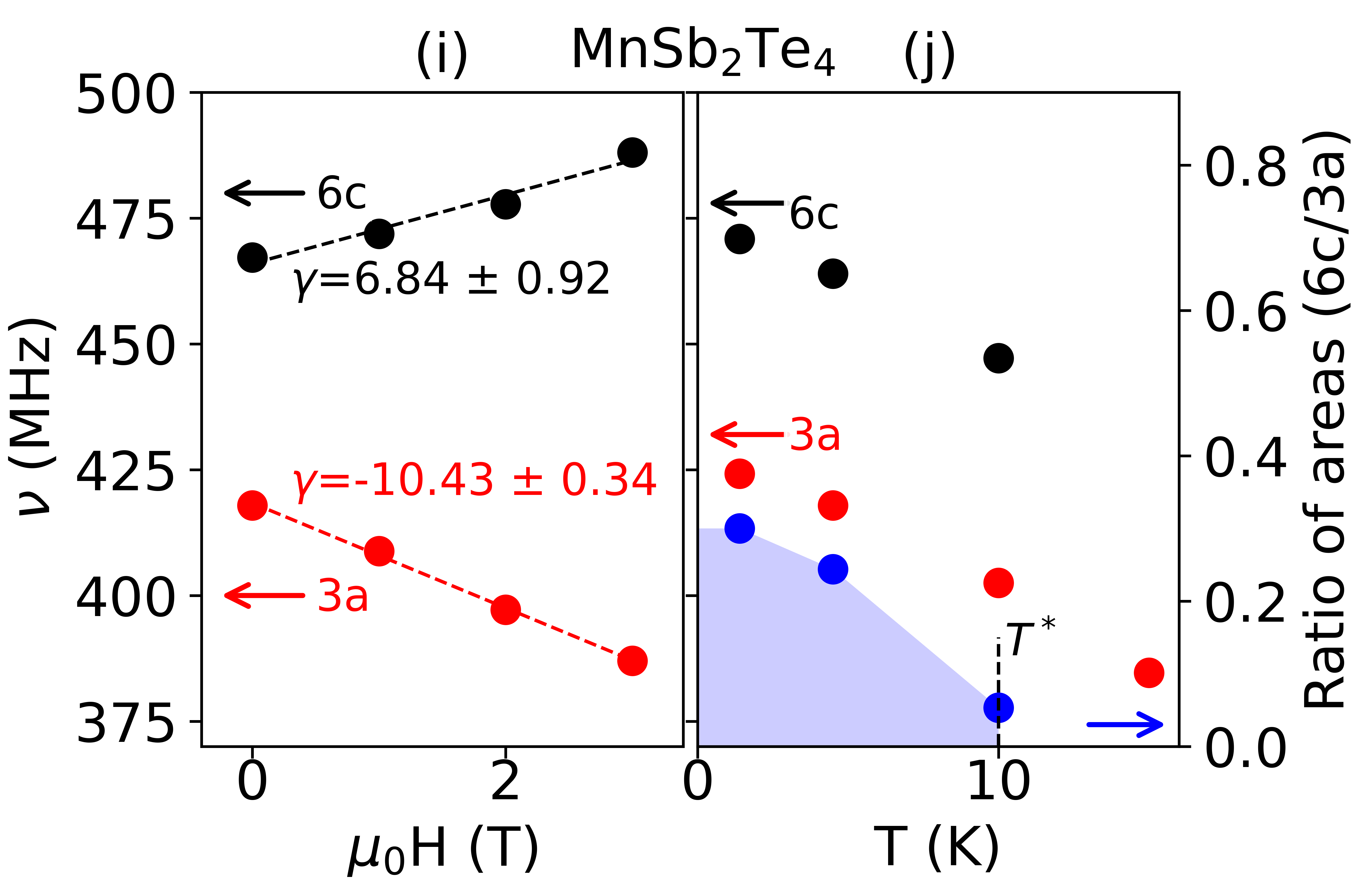}
  \label{fig:sfig3t}
\end{subfigure}%

\caption{(a-c), \Mn\ NMR spectra of \MBT~($\beta$ sample), \MBTw\ and \MST~respectively, at $T=1.4$ K in increasing applied fields, starting from ZF; (d-f) polycrystal vector composition  ${^{55}\mathbf{B}}=\mu_0\mathbf H+\mathbf{B}_{\mathrm{hf}}$ and their resulting spectral shifts for three simple cases: soft ferromagnet (FM, d), soft ferrimagnet, minority spin ($\downarrow$FIM, e), antiferromagnet (AFM, f); (g,h) field dependence of the \Mna\ and \Mnc\ mean frequency peaks for the \MBTnn~family; (i,j) \Mna\ and \Mnc\ peaks for \MST, field dependence (i) and temperature dependence (j) of their frequencies (black, red dots), temperature dependence of the ratio of their areas (blue dots, j).}
\label{fig:NMR}
\end{figure*}

\subsection{NMR peak assignment}

A 3D view of the spectra for polycrystalline samples of \MBT, \MBTw~and \MST \ at $\mathrm{T}$ = 1.4~K, in the frequency range 350 to 500 MHz, is shown in Fig.~\ref{fig:NMR}a-c (\MBTh~ in the Supplemental Material, Fig.~S.2) \cite{SM})  with ZF at the back and increasing fields \B\ towards the front. They all show two broad peaks patterns, each centered at a distinct frequency: $\nu_{\mathrm{a,c}}= {^{55}\gamma}\, |^{55}\mathbf{B}_{\mathrm{a,c}}|$. Therefore two distinct Mn sites experience a different total local field modulus \BMn$_{\mathrm{a,c}}$, i.e.~different values of \Bhf, as it would be expected for the main site \Mna\ and the anti-site \Mnc. The 5-10\% breadth of the frequency peaks is due to disorder in their vicinity, producing small variations of the local electronic environment, reflected in \Bhf. 

The relative area under the two peaks at 3T assigns the lower-frequency, majority peak to \Mna\ and the minority peak to \Mnc\ (Fig.~\ref{fig:NMR}a-c, the proportionality of the signal amplitude with the number of nuclei may not be guaranteed in ZF but it is recovered in 3T, see Sec.~\ref{sec:NMRmethods}). The frequencies confirm the assignment, since the six non-negligible transferred couplings $\cal B$ of \Mna\ and the three nearly vanishing ones for \Mnc~are both opposite to the on-site coupling ${\cal A}$. This simple argument is confirmed by DFT simulations of the hyperfine coupling at the Mn sites (Fig.~S.3).

\subsection{Antisite spin alignment}

The same 3D plots (Fig.~\ref{fig:NMR} a-c) show that the frequency splitting increases with the applied field. This can be understood from the field vector composition ${^{55}\mathbf{B}}=\mu_0\mathbf{H}+\mathbf{B}_{\mathrm{hf}}$. In simple cases, the relative local orientation of the \Mna\ and \Mnc\ magnetic moments is easily inferred from this vector composition, as shown in Fig.~\ref{fig:NMR} (d-f)
\begin{equation}
\label{eq:NMRfrequency}
^{55}\nu(H) =\nu_{\mathrm{hf}} 
\begin{cases}

 -\,{^{55}\gamma}\,\mu_0H & \text{FM},\,\uparrow\text{FIM}\\
 + \,{^{55}\gamma}\,\mu_0H & \text{$\downarrow$ FIM,}\\
  + \,\mathcal{O}\left(\left[\frac{\mu_0H}{B_{\mathrm{hf}}}\right]^2\right)& \text{AFM}
\end{cases}
\end{equation}
Namely, in the FM case, or equivalently, for the large magnetization sublattice of a ferrimagnet ($\uparrow$ FIM)  the moments align along the applied field, hence  $\mathbf{B}_{\mathrm{hf}}$ is opposite to $\mathbf{H}$ and the frequency decreases with increasing field. On the other hand, they anti-align to $\mathbf H$ for the small magnetization sublattice of a ferrimagnet ($\downarrow$ FIM), so there the frequency grows with $\mathbf{B}_{\mathrm{hf}}\parallel\mathbf{H}$. Finally, in the collinear AFM case, they do not align, and the superposition of all relative vector orientations in our polycrystalline samples leads to a broadening with no first order shift. \cite{SM}

Figures \ref{fig:NMR} g,h show the shifts vs. applied field for the \MBTnn~samples. The majority \Mna\ peak of \MBT\ (red stars) does not shift with $H$ up to 1 T, following the AFM behavior predicted by Eq.~\ref{eq:NMRfrequency}. Its small shift for 2 T$\le H\le 3$ T is consistent \cite{SM} with a canted antiferromagnetic (CAFM) state \cite{lee2019spin,bac2022topological,sass2020magnetic}. In contrast, \MBTw~and \MBTh~(black and blue symbols, respectively) follow the FM case, like the \MST~majority site, shown by the red symbols in Fig.~\ref{fig:NMR}-i, all fitted with negative slopes equal to $\gamma = -{^{55}\gamma}$ within error bars.

Neutron diffraction measurements on \MST\, have revealed the antiparallel alignment of the \Mna\ and \Mnc\ local moments. \cite{PhysRevX.11.021033, Riberolles2021} In our NMR data, this spin alignment is directly demonstrated in Fig.~\ref{fig:NMR}-i by the positive slope of $^{55}\nu(H)$ for the  \Mnc\ sites. Moreover,
we see exactly the same behavior for all \MBTnn\ materials (Fig.~\ref{fig:NMR} g,h), i.e., their septuple layers show the same FIM structure as \MST. While indications of this behavior in \MBT\ were previously seen in high-field magnetometry \cite{Lai2021}, the presented NMR results provide direct local evidence of the opposite spin alignment of \Mna\ and \Mnc\ for \emph{all} \MBTnn\ with $n=0, 1$ and 2. It is likely that the $n>2$ members of the \MBTnn\ family \cite{Amiraslanov.prb2022, Hu.sciadv2020, Lu.prx2021} should also display this FIM structure, pretty much as their $n=0-2$ analogs. Note that the $^{55}\nu(H)$ slope for \MBTnn\ in Fig.~\ref{fig:NMR}-h is slightly reduced \cite{SM} by an average canting angle $\theta$ between the sublattice magnetization and $\mathbf{H}$, according to  $\gamma = {^{55}\gamma}\cos\theta$.
This angle is small for the FM-like materials, but quite sizable for CAFM \MBT, reflecting the large powder average angle between the magnetization of its canted \Mna\ sublattices and the field. 

\subsection{Antisite moment temperature dependence}  The \Mn~NMR signal quickly disappears when increasing the temperature above $T=1.4$ K, as the NMR $T_2$ relaxation time gets shorter than the instrumental dead-time. For \MST~the \Mna~ZF NMR frequency (red symbols in Fig.~\ref{fig:NMR} j) is detected up to $t = T/T_{\mathrm{m}} \sim 0.6$, whereas the \Mnc~ZF NMR frequency (black symbols) is detected only up to $t = 0.4$. They both decrease towards the second order transition at \Tm, following the order parameter. However, the ratio of their peak areas (blue symbol) decreases much more quickly over the same range, vanishing around  $t=0.4$. This implies that the \Mnc\ peak disappears in a first-order-like fashion at a temperature \Tstar\ well below \Tm. 

Summarizing, NMR at 1.4 K directly detects the opposite alignment of the \Mna~and \Mnc~sublattices in all of the systems studied here, and it demonstrates that in the Sb-based compound the antisite \Mnc~disorders above \Tstar, well below \Tm. The same information is not accessible for \MBTnn, due to a combination of lower ordering temperatures and shorter $T_2$.   

\section{$\mu$SR results}
\label{sec:muSR}
\begin{figure*} 
\includegraphics[width=0.975\columnwidth]{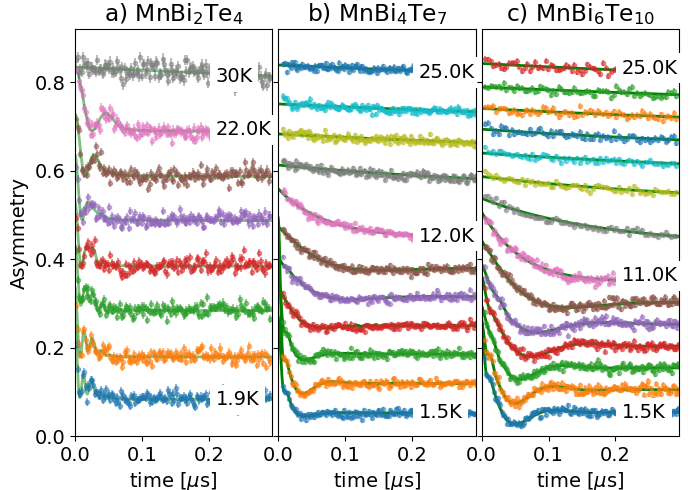}\quad
\includegraphics[width=1\columnwidth]{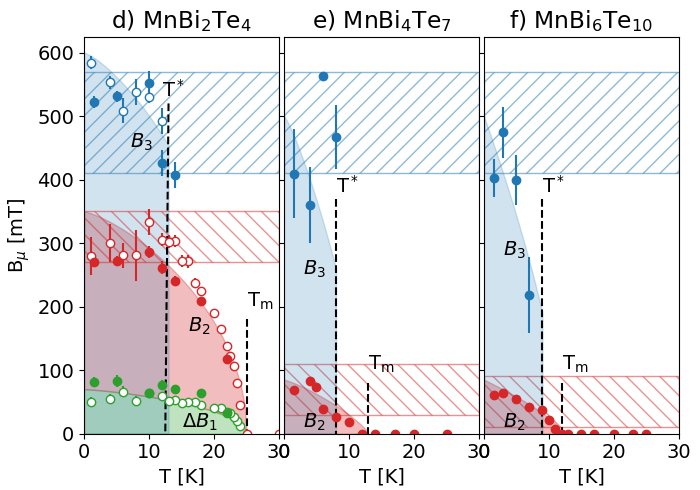}\quad
\includegraphics[width=1.1\columnwidth]{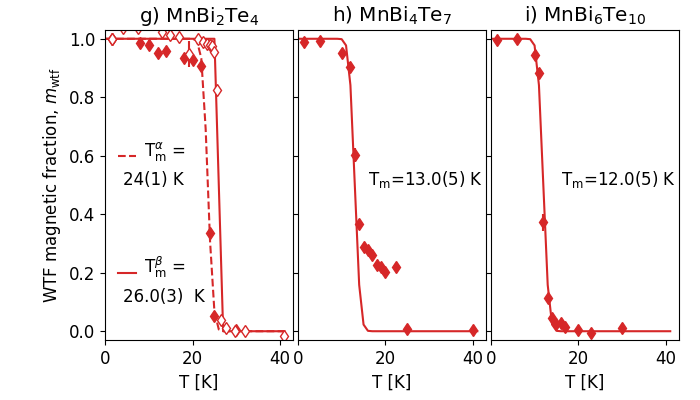}\quad
\includegraphics[width=0.9\columnwidth]{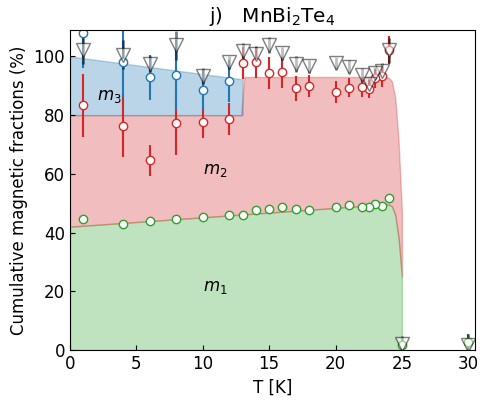}
\caption{\muSR\ in \MBTnn, $n=0,1,2$. (a-c) Early time ZF asymmetries at various temperatures with best fits (solid curve), displaced vertically for clarity; (d-f) Temperature dependence of the internal fields $\Delta B_1, B_2, B_3$ (green, red, blue symbols, respectively), the hatched bands show
the values predicted by DFT at $T=0$, 
shaded areas are guides to the eye (Eq.~\ref{eq:B(T)});
(g-i) WTF magnetic volume fraction \fwtf~vs.~temperature;  (j) ZF transverse cumulative magnetic fractions $m_1$ (green circles),  $m_1+m_2$ (red  circles) and $m_1+m_2+m_3$ (blue  circles), plus longitudinal magnetic fraction \fl~(grey triangles), cfr. Eq.~\ref{eq:ZFmuSR}. Data in panel (a) and  open symbols in panel (d,g,j) correspond to the \MBT\ $\alpha$ sample.}
\label{fig:muSR}
\end{figure*}
Spin-polarized muons implanted in polycrystalline samples stop in few lowest energy interstitials. In ZF and for $T <T_{\mathrm{m}}$ the muon spin precesses around the local magnetic field \Bmu, due to ordered moments, producing coherent oscillations at a frequency $\gamma_\mu B_\mu$ in the asymmetry of the muon decay at early times ($\gamma_\mu=135.554$ MHz/T, see Sec.~\ref{sec:methods}). 

Selected ZF-\muSR\ early-time asymmetries $A(t)$ are shown in Fig.~\ref{fig:muSR} a-c for the \MBTnn~materials at different temperatures,  with their best fit to a minimal model, Eq.~\ref{eq:ZFmuSR} in Sec.~\ref{sec:methods}. The model describes the internal field distribution \pB\ probed by muons as a few Gaussian components of very broad width $\Delta B_i$. Starting from \MBT~(Fig.~\ref{fig:muSR} a), the asymmetry shows a fast, overdamped initial relaxation  and a second damped oscillation below \Tm. Both components correspond to appreciable internal fields, the fast initial Gaussian decay to a mean value smaller than its width $0\lesssim B_1<\Delta B_1$, and the visible oscillation to an observable mean value $B_2>\Delta B_2$. At low temperature a third oscillating component ($B_3>\Delta B_3$) appears. By comparison, the $n=1,2$ members differ from $n=0$ in {\em i)} the absence of the $B_1<\Delta B_1$ fast initial decay and {\em ii)} a lower field $B_2$ value, whereas, they also display {\em iii)} a  high field $B_3$ component, that sets in only at lower temperatures. The presence of two distinct oscillations, both heavily damped, is more evident in the  low temperature best fits of Fig.~S.4. \cite{SM}

\subsection{Magnetic transitions}
The temperature dependence of the internal fields, $\Delta B_1(T)$ (the width of that distribution), $B_2(T)$  and $B_3(T)$, are shown in  Fig.~\ref{fig:muSR} d-f. It reveals two common features among all three family members: $\Delta B_1$ and $B_2$ correspond to the order parameter and vanish at the second-order magnetic transitions, \Tm; in contrast,  $B_3$ vanishes abruptly at \Tstar, inside the ordered phase, without any corresponding anomaly in $\Delta B_1$, $B_2$.

Weak transverse field (WTF) $\mu$SR provides the amplitude of the spin precession in a small applied field ($\mu_0 H \ll \Delta B_1, B_2, B_3$). Below \Tm~this amplitude drops abruptly.  The \MBTnn~{\em magnetic volume fraction} \fwtf~shown in Fig.~\ref{fig:muSR} g-i is obtained from WTF measurements (Eq.~\ref{eq:mtf} in Sec.~\ref{sec:methods}) and demonstrates that \MBT~and \MBTh~undergo very sharp transitions, at \Tm,  (the width of the transition for a 90\%-10\% volume reduction is $\Delta T<1$~K), despite their relatively large atomic disorder implied by the presence of antisites. The $n=1$ sample displays a sharp transition as well, but a 20\% contribution, which we attribute to intergrowths of the \MBT\ phase, is also visible (Fig.~\ref{fig:muSR} h). 

Similar results were obtained for \MST. \cite{sahoo2023impact} Notably, we measured two distinct \MBT\ samples, labeled $\alpha$ and $\beta$ (open and filled symbols, respectively, in Fig.~\ref{fig:muSR} d,g,j), which readily show distinguishable transitions (\Tm =  26.0(3), 24(1) K), due to slightly different preparation conditions. 
The residual WTF amplitude well below \Tm~ is due to a small fraction of muons implanted outside the sample, that {\em do not} experience its spontaneous internal magnetic field distribution.

Separate magnetic volume fractions $m_{i}$ are also derived from the ZF normalized amplitudes at each internal field $B_i$ (Eq.~\ref{eq:mt}, Sec.~\ref{sec:MuSR methods}). Independently, the total volume fraction \fl~is obtained from the longitudinal amplitude (Eq.~\ref{eq:ml}). Figure \ref{fig:muSR} j shows the cumulative sum of the three transverse fractions for \MBT-$\alpha$. Both \fl~(grey symbols) and $m_{1}+m_{2}$ (red symbols) drop sharply at the transition $\mathrm{T_m}$, in agreement with \fwtf. In contrast, $m_{3}$ disappears abruptly at \Tstar=12(1) K, suggesting that this component originates from muon sites sensitive to the subtle change that takes place across that point. This is reminiscent of our NMR \MST~results. For the \MBTnn~samples, the temperature coincides with a clear anomaly appearing in magnetization  (Fig.~S.1 and Tab.~S.I), suggesting that \Mnc~moment are undergoing fast (paramagnetic-like) reorientations above \Tstar. Further insight crucially requires the correct identification of the muon-stopping sites.
 
\subsection{\label{sec:MuonAntisites}The antisites from the \muSR~perspective}

\begin{figure}[!htbp]
\centering

\includegraphics[width = 0.8\linewidth]{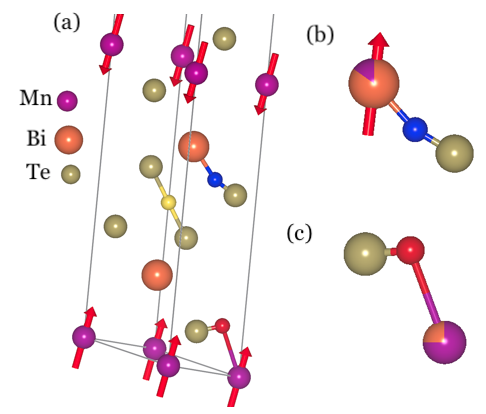}
    \caption{Representative muon sites as small colored atoms in \MBT~(and \MST) by \DFTm: \TM~(red), \TB~(blue atom), \TT~(yellow). (a) Half the primitive cell with \Mna~only, (b) \TBM~ and c) \TMB.} 
    \label{fig:sites}
\end{figure}

The muon sites were identified and their respective $T=0$~K field values, $B_\mu$, were computed using a standard protocol, known as \DFTm, \cite{moeller2013,bonfa2016,onuorah2018,bonfa2018}. These are explained briefly in Sec.~\ref{sec:DFT methods} and discussed in more details in the Supplemental Material. \cite{SM} In the ideal \MBT\ crystal, this protocol identifies three stable sites, shown in Fig.~\ref{fig:sites} as \TM~(red sphere), \TB~(blue sphere),  and \TT~(yellow sphere) and reported in the top box of Tab.~\ref{tab:sites}. The mean field values at these sites agree nicely with the experimental component $B_1$ (\TB\ and \TT) and $B_2$ (\TM), respectively.  Notice that the correspondence between sites and fields is not bijective (more sites may contribute to the same field distribution) and that the uncertainty in the \DFTm\ derived values is below 25\%.

\begin{table}[!ht]
    \caption{$T=0$ local field at the representative muon sites in \MBT, identified by color as in Fig.~\ref{fig:sites} (see Supplemental for details \cite{SM}). Boxes refer to muons sites: top, far from antisites; bottom, nn to an antisite. Colored dashes refer to colors of symbols in Fig.~\ref{fig:muSR} d.} 
    \label{tab:sites}
    \begin{tabular}{| l |  c | c | c | c |}
    \hline
        & nn & \DFTm & \multicolumn{2}{c|}{ Experiment  } \\
        Muon site label& antisite &  \Bmu (mT) & $B_i$ (mT) & term\\
        \hline
             {\LARGE \color{blue} $\bullet$} \TB &-&   93  &\multirow{2}{*}{$0(80)$} & {\multirow{2}{*} {{\color{Green} $\boldsymbol{-}$} $\Delta B_1$ }}\\
         {\color{Goldenrod}\LARGE$\bullet$} 
         \TT &  -&  0 & &\\

        {\LARGE \color{Red} $\bullet$} \TM  & -&  314   &  290(10) & {\color{Red} $\boldsymbol{-}$} $B_2$\\
        \hline
        {\LARGE \color{blue} $\bullet$} \TBM  & \Mnc&  527   &  550(30) & {\color{NavyBlue} $\boldsymbol{-}$} $B_3$ \\
        {\LARGE \color{red} $\bullet$} \TMB &\Bia &   95  &  $0(80)$ &   {\color{Green} $\boldsymbol{-}$} $\Delta B_1$ \\
        \hline
    \end{tabular}
\end{table}
 
The presence of intermixing modifies these findings in two ways: (i) inherent disorder broadens considerably all field distributions, producing large widths $\Delta B_i$, and (ii) the extra moment modifies significantly the mean field values at muon sites nearest neighbor (nn) to \Mnc~and \Bia. We label these two modified sites as \TBM~and \TMB, respectively, and report their properties in the lower box of Tab.~\ref{tab:sites}.  Both these consequences are observed experimentally, in particular the calculation for the muon site \TBM~agrees with local field $B_3$, while \TMB~contributes to $B_1$.

Let us now turn to the \MBTnn~compounds with $n=1, 2$, where the sites closest to the \Mna~layer are predicted to be very similar to those of $n=0$. The farther high symmetry \TT~site, instead, is replaced by more than one site in the intervening quintuple layers. Since all of them are far removed from \Mna, we expect a large majority of muon sites characterized by very small or vanishing local field values. However, $n=1,2$ samples show non-vanishing net magnetic moment , hence \Bmu~has an additional Lorentz field term 
$B_{\mathrm{L}} = 4\pi M/3\approx 70,50$ mT, respectively. This contributes negligibly to the high field of the \TBM\ site, $B_3\gg$~\BL, but significantly to the low field \TMB, \TB~and \TT\ sites. 
This justifies both the second experimental field value $B_2\approx B_{\mathrm  L}$ (Fig.~\ref{fig:muSR} e,f), its large fraction $f_2$ and the disappearance of the $f_1, B_1=0$ signal.

The agreement of all these predictions, shown as hatched color bands in Fig.~\ref{fig:muSR} d-f, with $T\rightarrow 0$~K experimental data is altogether remarkable. It is the more so, in as much the same three DFT muon sites support a coherent, simple interpretation of the data for up to three experimental fit components, in four different compounds, over two magnetic phases.

\section{Discussion and conclusions}
{\label{sec:conclusions}}

An important result from NMR is that the size of the moment on Mn obtained from the hyperfine field, in first approximation, is proportional to the on-site coupling ${\cal A}$, which is roughly 10~T/$\muB$ for all 3d ions and for \Mn~in particular. \cite{Freeman1965,Kubo1969,Allodi2002}
An estimate of this coupling comes from the ZF NMR frequency of \Mna\ and \Mnc. Taking the value of the latter for its much smaller transferred terms $\cal B$, and, conservatively, half their difference as the uncertainty, we get roughly the same moment for all three \MBTnn~samples, $\mu_{\mathrm {Mn}}=4.3(2)\,\muB$, in agreement with neutron diffraction \cite{Ding2020}.

Importantly, fast \Mnc~spin fluctuations above \Tstar~ justify the disappearance of the $f_3$ component, assigned consistently to the \TBM~site and mostly due to the nn \Mnc~moment. This observation is common to all three compositions. We recall that the same conclusion is drawn from the NMR findings on \MST~(Sec.~\ref{sec:NMR}), confirmed by \muSR~as well \cite{SM}.  In addition, \muSR~results confirm the first order character of \Tstar, since the internal field $B_3$, proportional to the \Mnc~moment, is still large when the signal amplitude vanishes (Fig.~\ref{fig:muSR} d-f, mimicking the NMR results of Fig.~\ref{fig:NMR} j).

The anomaly at \Tstar~is confirmed by magnetization data, \cite{SM} which, however, are obtained in an applied field. 
Recall that in \MBTh~the local field $B_2$ is assigned to the low field sites \TB, \TT~and \TMB, \cite{SM}, in the presence of an additional comparable Lorentz field, \BL, due to the net domain magnetization. The smooth behavior of $B_2$ across \Tstar~indicates that \BL~survives also in zero applied field above the transition, suggesting that a dominant FM stacking of \Mna~layers persists above \Tstar~ in zero field, therefore it is not induced by the external field. 

The orientation of the antisite static moments successfully shows the sign of their dominant local exchange, displaying a universal behavior in this family, but this does not tell us the full magnetic structure of each material, that varies in the family.  Actually, for \MBT~ we can assume that \Tm~is a N\'eel transition and the sample is AFM in zero field. Here, \Tstar\ (detected in ZF) clearly merges with the low temperature transition seen in $M(T)$ (Fig.S1 \cite{SM}). For \MBTw~the presence of a cusp in $M(T)$ suggests AFM bulk behavior at \Tm, but \muSR~shows the presence of an $n=0$ contribution and we avoid further considerations. Finally, \MBTh~ was already shown to be FM under certain growth conditions. \cite{tcakaev2023intermixing,Yan.nl2022}

\begin{figure}[!htbp]
\centering
\includegraphics[width = 0.8\linewidth]{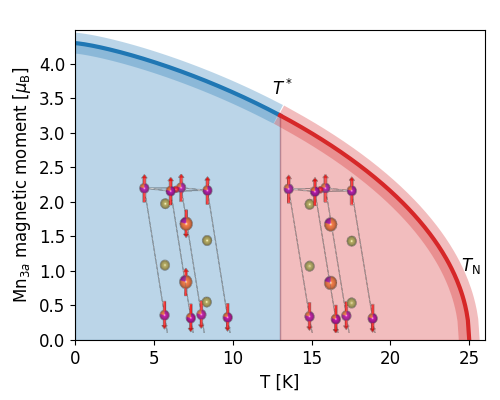}
    \caption{\MBT~phase diagram summary: temperature dependence of the \Mna~magnetic moment, rescaled from the \muSR~fields $\Delta B_1(T), B_2(T)$, and $T\rightarrow 0$ value obtained from ZF \Mna~NMR (shaded bands translate uncertainty in the latter). Insets: magnetic structures in one septuple layer of the primitive cell.} 
    \label{fig:phasediagram}
\end{figure}

Figure \ref{fig:phasediagram} summarizes our findings in a schematic phase diagram for the A-type antiferromagnet \MBT, where the $T\rightarrow 0$ magnitude of the static Mn moment is taken from NMR and its temperature behavior from the interpolation by Eq.~\ref{eq:B(T)}  on the ZF \muSR~fields $\Delta B_1(T), B_2(T)$, of Fig.~\ref{fig:muSR} d. The blue region is characterized by the static ordering of the diluted \Mnc~moments, aligned antiparallel to the main \Mna~moments, as NMR demonstrates. The static moment at \Mnc~vanishes in the red region.

A similar plot for \MBTh~is reported in the Supplemental Material \cite{SM} (\MBTw~data are less reliable due to the presence of a sizable fraction of \MBT~layer intergrowths, Fig.~\ref{fig:muSR} h).  Remarkably, in all three compositions, the mean \Mna~order does not change appreciably across \Tstar. 

The red-shaded high-temperature phase is characterized by a very sharp second-order transition, as it is witnessed in all samples by the abrupt vanishing of the magnetic volume fraction, both by WTF and ZF \muSR~(Fig.~\ref{fig:muSR} g-i). We highlight the unique ability of \muSR, as opposed to both magnetometry and neutron scattering, to distinguish the reduction of the magnetic moment, encoded in the local field, from the volume fraction, encoded in the amplitude of the signal. Local disorder, like that expected from magnetically coupled, random antisites, could yield a distribution of transition temperatures, and hence a more progressive reduction of the volume fraction than that displayed in all our samples.  This suggests that, from a magnetic point of view, the high-temperature ordered phase approaches closely the ideal, intermixing-free material. This provides both the antiferromagnetic and the ferromagnetic versions of a close-to-ideal magnetic topological insulator. 

Early sublattice decoupling is well documented, for instance in intermetallic compounds, \cite{Cugini2022, Willis2001,Damay2020, Garlea2019,cadogan2001,Morrow2013,Lyubutin1992,Orlandi2016} and it often involves exchange coupling frustration, which may also play a role in the present case \cite{li2021glassy}. Since intermixing is a common feature for all materials, including cation species with similar radii, we speculate that early antisite disordering may take place more often than one thinks. Therefore, our findings  may have a more general impact than just on the present materials. 

\section{Methods}
\label{sec:methods}

\subsection{Synthesis}
Polycrystalline samples of \MBTnn~with $n=0, 1,2$ and of  \MST~were prepared by high-temperature solid-state reactions, kept as non-compacted powders for NMR experiments, and pressed into pellets for the \muSR\ experiments. Two \MBT  \, (denoted $\alpha$ and $\beta$) powdered samples were obtained by two distinct high-temperature annealing routes.

The \MBT \, $\alpha$-sample was prepared from a mixture of pre-synthesized MnTe and Bi$_2$Te$_3$ taken in the ratio 0.87:1.05. The powders were handled in an argon-filled dry glovebox (MBraun), homogenized in a dry ball-mill (Retsch, MM400) at 20~Hz for 20~min and then pressed into a 6-mm pellet (2 tons, 30 sec). The pressling was placed into a quartz ampoule, sealed off under dynamic vacuum ($3 \times 10^{-3}$~mbar) and annealed in a temperature-controlled tube furnace (Reetz GmbH) following the procedure developed in~\cite{Zeugner.cm2019}. 

The MnBi$_4$Te$_{7}$ and MnBi$_6$Te$_{10}$ samples were obtained from a corresponding, stoichiometric  mixture of the powdered binaries which were handled similar to the ones above. The former was annealed at 585\degree C for 10~days (heating rate 1\degree / h) and the latter was annealed at 575\degree C for 4~days (heating rate 90\degree / h). Both samples were water-quenched. 

 The polycrystalline \MBT \, $\beta$-sample was synthesized by co-melting of pre-synthesized MnTe and \BiTe\ in an evacuated quartz ampoule at a temperature of about 980 $^\circ$C for 12 h, followed by slow cooling down to 580 $^\circ$C at the rate of 5 $^\circ$/h. This temperature was kept for 12 h followed by air-quenching. Then, the polycrystalline sample was ground as a fine powder and converted into a pellet and was then sealed in a quartz container under the pressure of 10$^{-4}$ Pa. The ampoule was further heated up to 585 $^\circ$C for 8 h, then kept at this temperature for about 240 h followed by air-quenching. The grinding and annealing process was repeated twice to achieve a homogenized phase-pure compound.  

The MnSb$_2$Te$_4$ sample was synthesized from a  stoichiometric mixture of the elements (Sigma-Aldrich, 9N5, Mn reduced prior to synthesis) that was ball-milled at 20~Hz for 20~min, pelletized and annealed at 550\degree C for 8~days, and finally quenched.

 \subsection{X-ray diffraction and energy-dispersive X-ray spectroscopy}

Phase purity of all samples but $\beta$ was  analyzed by powder X-ray diffraction (PXRD) on a Malvern Panalytical Empyrean 3 diffractometer, employing Cu$K\alpha_1$ radiation ($\lambda = 1.54059 \AA$) and set in Bragg-Brentano geometry. Lattice parameters refinement was conducted by Le Bail method to ensure the correct assignment of the \MBTnn \, phases. A PXRD pattern of the $\beta$-sample was taken on a Bruker D2 PHASER diffractometer using Cu$K\alpha_1$ radiation within the scanning range of $2\theta=5-75$. All \MBTnn \,samples were found to be single-phase by comparing with the published results~\cite{Zeugner.cm2019, Aliev.jac2019, Souchay.jmcc2019, vidal2019topological, tcakaev2023intermixing,souchay2019layered}, without admixtures of other binary or ternary phases. 

The \MST \, sample contained a very small imprurity of MnTe$_2$ assessed as 2~wt.\% by Rietveld refinement. The lattice constants of the main \MST \, phase were refined as $a = 4.2416(1), c = 40.883(2)$ within the $R\bar{3}m$ space group. This material can flexibly accommodate strong Mn/Sb intermixing~\cite{folkers2022occupancy, liu2021sitemixing, SAHOO2023101265} and, therefore, adopt various total compositions. To achieve the highest accuracy in the determination of the chemical composition of our  powdered sample, we conducted  calibrated EDX measurements. For that, parts of the sample were cast into synthetic resin (versosit)  pucks, sputtered with a gold layer and painted with conductive silver paint to avoid charge accumulation.  EDX spectra were recorded with a high-resolution SEM
EVOMA 15 (Zeiss) equipped with a Peltier-cooled Si(Li) detector
(Oxford Instruments) employing 30 kV acceleration voltage. Element
quantification was obtained from least-square fitting of edge models
(Mn-K, Te-L, Sb-L) invoking k-factor calibration from the
stoichiometric samples of similar composition (Sb$_2$Te$_3$ and MnTe).
To assess systematic errors stemming from the different edge and
reference choices, we included Sb$_2$Te$_3$
and MnTe references for Te in our quantification statistics. The determined  composition of the sample was Mn$_{0.87(1)}$Sb$_{2.03(1)}$Te$_{4.00(1)}$ which is in line with its magnetic properties reported in~\cite{sahoo2023impact}.

\subsection{NMR}
\label{sec:NMRmethods}
The NMR spectra were measured in a He-flow cryostat by means of the HyReSpect home-built phase coherent broadband spectrometer \cite{Allodi2005}. 
Spin-echoes were excited at discrete frequency points by refocusing P--$\tau$--P Hahn  
radio-frequency (rf) pulse sequences, with optimized pulse duration and intensity to maximize the resonance signal, and shortest $\tau$ delay (limited by the apparatus dead time of few $\mu$s). 

Spectra are reconstructed from the maximum of the spin-echo Fourier transform amplitude at each frequency, corrected for the frequency-dependent sensitivity and nuclear Boltzmann factors. The normalized values correspond to the spectral distribution of hyperfine fields at the $^{55}$Mn nuclei. When the best fit of the two spectra ($\zeta$ = \Mna, \Mnc) require more than one Gaussian component each, their mean frequency is calculated from the corresponding weights $A_{\alpha,i}$ as the first moment $ \sum_i A_{\zeta i}\nu_{\zeta i}/\sum_k A_{\zeta.k}$.

ZF nuclear echoes were collected with a non-resonant probe circuit, by virtue of the large rf enhancement $\eta=H_1^*/H_1$, where $H_1$ is the applied field at the radio frequency $\omega$, and $H_1^*$ is the $\omega$ oscillating component of the huge hyperfine field \Bhf, following the electronic moment
response to $H_1$.\cite{Meny2021,Sidorenko2006} The factor $\eta$ is very sensitive to nanoscopic and mesoscopic changes in the nucleus environment and this does not guarantee a uniform proportionality between spectral area and number of resonating nuclei. Relative proportionality is recovered in the more uniform resonant conditions obtained at high static applied fields. 

\subsection{\muSR}
\label{sec:MuSR methods}

The $\mu$SR experiments were carried out at the Paul Scherrer
Institute, Villigen, Switzerland, on the GPS spectrometer.

A minimal choice for the best-fit function of the time domain ZF asymmetry, arising from parity violation in the weak muon decay, is the following

\begin{align}
\label{eq:ZFmuSR}
A_{ZF}(t) & =  A_0\left[ f_{\mathrm{T}1}\, e^{-\sigma_1^2t^2/2} 
+\sum_{i=2}^{3} f_{\mathrm{T}i}\, \cos(\gamma_{\mu}B{_1}t) \,e^{-\sigma_i^2t^2/2}\right. \nonumber \\ & +   \left.f_{\mathrm{L}}\,e^{-t/T_1}\right],
\end{align}
distinguishing the fast relaxing, precessing fractions $f_{\mathrm{T}}= \sum_{i=1}^3 f_{\mathrm{T}i}$ ($\mathrm{T}$ for transverse with respect to the initial muon spin direction) from the slow relaxing fraction $f_{\mathrm{L}}$ ($\mathrm{L}$ for longitudinal), which  corresponds to local field components parallel to the initial muon spin direction. The precession relaxation rates are due to the width of each field distribution, $\sigma_i=2\pi \gamma_\mu\Delta B_i$, their fractions obey $f_{\mathrm{T}}+f_{\mathrm{L}}=1$ and the maximum experimental asymmetry, $A_0$, is calibrated at high temperature. Polycrystalline averaging leads to $2f_{\mathrm{L}}=f_{\mathrm{T}}$, but a very fast transverse decay and a small fraction of muons stopping outside the sample may relax this ideal condition. 

The WFT time domain signals are best fitted by an oscillatory and two relaxing functions
\begin{eqnarray}
\label{eq:WFT}
       A_{TF}(t) &=& A_0 \left[{f_p}\,\cos(\gamma B +\phi)\, e^{-\lambda_p t}\right.\nonumber\\
       && \left.+f_{\mathrm{T}}\,e^{\sigma_{\mathrm{T}}^2t^2/2}  + f_{\mathrm{L}}\, e^{-\lambda_{\mathbf{L}} t}\right]  
\end{eqnarray}
where $\phi$ is an initial phase and $f_p$ is the muon fraction that does not experience strong local hyperfine fields, which includes muons stopping outside the sample and inside paramagnetic microdomains. The former corresponds to the low temperature residual value $f_{p0}$, whereas for $T\ll$ \Tm, one has $f_p=1, f_{\mathrm{T}}=f_{\mathrm{L}}=0$. The two relaxing function represent the transverse and longitudinal fractions of muons experiencing internal fields.

Three independent determinations of the magnetic volume fraction are given by
\begin{eqnarray}
       m_{\mathrm{wtf}}(t) &=& \frac{f_p(T)-f_{p0}}{1-f_{p0}},\label{eq:mtf}\\
   m_{\mathrm{T}} &=& \sum_{i=1}^3 m_i\qquad m_i = \frac {f_{\mathrm{T}i}}{f_{t0}}, \label{eq:mt}\\
       m_{\mathrm{L}} &=& = \frac 3 2 (1-f_{\mathrm{L}}(T)),\label{eq:ml}
\end{eqnarray}
where $f_{\mathrm{T}0}=\lim_{T\rightarrow 0} f_{\mathrm{T}}(T)$ and Eq.~\ref{eq:mt} neglects $f_{p0}$, since in ZF $f_{\mathrm{L}}$ is indistinguishable from $f_p$. They are used in Fig.~\ref{fig:muSR} g.h (Eq.~\ref{eq:mtf}) and j (Eq.~\ref{eq:mt} colored and gray symbols, for Eq.~\ref{eq:mt} and \ref{eq:ml}, respectively).

Lastly, the internal fields $\Delta B_1 = \sigma_1/2\pi\gamma_\mu$, and $B_2$ proportional to the order parameter are fitted to a standard phenomenological function \cite{Blundell}

\begin{equation}\label{eq:B(T)}
b(T) = \left[1-\left(\frac T {T_{\mathrm{m}}}\right)^\gamma\right]^\delta,
\end{equation}

used both in the shaded colors of Fig.~\ref{fig:muSR} d-f and in Figs.~\ref{fig:phasediagram}, S.8.

\subsection{DFT}%
\label{sec:DFT methods}

The DFT calculation protocols for determining the muon implantation sites  \cite{moeller2013} and the contact contribution to the hyperfine interactions in magnetic compounds~\cite{bonfa2016,onuorah2018}  are well established. State of the art DFT calculations are performed on a magnetic supercell, including an extra impurity hydrogen atom. Sampling of the starting impurity supercell coordinates and lattice relaxation by force minimization yield minimum total energy sites and their spin couplings, that allow the calculation of the local field \cite{bonfa2018}. Full calculation details are reported in the Supplemental Material \cite{SM}.

\section{Data availability}

The original data are available at \cite{muondata} \url{http://http://musruser.psi.ch}, instrument GPS, years 2020, 2021, title includes the chemical formula. NMR data, the DFT input and results are available at Materials Cloud Archive \cite{materialscloud} \href{https://archive.materialscloud.org/record/2024.20}{DOI:10.24435/materialscloud:he-f5}. 



\section{Acknowledgments}

This work was supported by the Deutsche Forschungsgemeinschaft (DFG) within the W\"urzburg-Dresden Cluster of Excellence on Complexity and Topology in Quantum Matter -- \textit{ct.qmat} (EXC 2147, project-id 390858490) and the SFB 1143 (project-id 247310070). IJO and RDR acknowledge financial support from PNRR MUR project ECS-00000033-ECOSISTER and also acknowledge computing resources provided
by the STFC scientific computing department’s SCARF cluster and CINECA under ISCRA Project ID  IsCa4. The cost of the HyReSpect equipment used for this investigation was partly supported by the University of Parma through the Scientific Instrumentation Upgrade Program 2020. This work is partially based on experiments performed at the Swiss Muon Source (S$\mu$S), Paul Scherrer Institute, Villigen, Switzerland. MMO acknowledges the support by MCIN/ AEI /10.13039/ 501100011033/ (Grant PID2022-138210NB-I00) and "ERDF A way of making Europe" as well as MCIN with funding from European Union NextGenerationEU (PRTR-C17.I1) promoted by the Government of Aragon. We acknowledge MSc.~Fabian~Lukas and MSc.~Laura Mengs (Technische Universität Dresden) for their contributions to inorganic synthesis as student assistants. We are grateful to the group of Prof.~A.~Lubk (TU Dresden, IFW Dresden) and to Mrs.~G.~Kreutzer for the calibrated EDX measurements and insightful discussions. 

\bibliography{references}

\end{document}


\title{Ubiquitous order-disorder transition in the Mn antisite sublattice of the \MBTnn\ magnetic topological insulators}

\author{M.~Sahoo}
\thanks{MS and IJO contributed equally}
\affiliation{Leibniz IFW Dresden, Helmholtzstra\ss{}e 20, D-01069 Dresden, Germany}
\affiliation{Institut f\"{u}r Festk\"{o}rper- und Materialphysik, Technische Universit\"{a}t Dresden, 01062 Dresden, Germany}
\affiliation{Würzburg-Dresden Cluster of Excellence ct.qmat, Germany}
\affiliation{Dipartimento di Scienze Matematiche, Fisiche e Informatiche, Universit\'a di Parma, Parco delle Scienze 7A, I-43124 Parma, Italy}

\author{I.J.~Onuorah}
\thanks{MS and IJO contributed equally}
\affiliation{Dipartimento di Scienze Matematiche, Fisiche e Informatiche, Universit\'a di Parma, Parco delle Scienze 7A, I-43124 Parma, Italy}

\author{L.C.~Folkers}
\affiliation{Institut f\"{u}r Festk\"{o}rper- und Materialphysik, Technische Universit\"{a}t Dresden, 01062 Dresden, Germany}
\affiliation{Würzburg-Dresden Cluster of Excellence ct.qmat, Germany}

\author{E.V.~Chulkov}
\affiliation{Donostia International Physics Center, 20018 Donostia-San Sebastián, Spain}
\affiliation{Departamento de Polímeros y Materiales Avanzados: Física, Química y Tecnología, Facultad de Ciencias Químicas, Universidad del País Vasco UPV/EHU, 20018 Donostia-San Sebastián, Spain}
\affiliation{Centro de Física de Materiales (CFM-MPC), Centro Mixto (CSIC-UPV/EHU), 20018 Donostia-San Sebastián, Spain}
\affiliation{Saint Petersburg State University, 199034 Saint Petersburg, Russia}
\author{M.M. Otrokov}
\affiliation{Instituto de Nanociencia y Materiales de Aragón (INMA), CSIC-Universidad de Zaragoza, 50009 Zaragoza, Spain}
\author{Z.S. Aliev}
\affiliation{ Baku State University, AZ1148 Baku, Azerbaijan}
\affiliation{ Institute of Physics Ministry of Science and Education Republic of Azerbaijan, AZ1143 Baku, Azerbaijan}
\author{I.R.~Amiraslanov}
\affiliation{ Institute of Physics Ministry of Science and Education Republic of Azerbaijan, AZ1143 Baku, Azerbaijan}
\author{A.U.B.~Wolter}
\affiliation{Leibniz IFW Dresden, Helmholtzstra\ss{}e 20, D-01069 Dresden, Germany}

\author{B.~Büchner}
\affiliation{Leibniz IFW Dresden, Helmholtzstra\ss{}e 20, D-01069 Dresden, Germany}
\affiliation{Institut f\"{u}r Festk\"{o}rper- und Materialphysik, Technische Universit\"{a}t Dresden, 01062 Dresden, Germany}
\affiliation{Würzburg-Dresden Cluster of Excellence ct.qmat, Germany}

\author{L.~T.~Corredor}
\affiliation{Leibniz IFW Dresden, Helmholtzstra\ss{}e 20, D-01069 Dresden, Germany}
\author{Chennan~Wang}
\address{Laboratory for Muon Spin Spectroscopy, Paul-Scherrer-Institute, CH-5232 Villigen PSI, Switzerland.
}

\author{Z.~Salman}
\address{Laboratory for Muon Spin Spectroscopy, Paul-Scherrer-Institute, CH-5232 Villigen PSI, Switzerland.
}

\author{A.~Isaeva}
\affiliation{Leibniz IFW Dresden, Helmholtzstra\ss{}e 20, D-01069 Dresden, Germany}
\affiliation{Van der Waals-Zeeman Institute, Department of Physics and Astronomy, University of Amsterdam, Science Park 094, 1098 XH Amsterdam, The Netherlands}

\author{Roberto De Renzi}
\email[Corresponding address: ]{roberto.derenzi@unipr.it}
\affiliation{Dipartimento di Scienze Matematiche, Fisiche e Informatiche, Universit\'a di Parma, Parco delle Scienze 7A, I-43124 Parma, Italy}

\author{G.~Allodi}
\affiliation{Dipartimento di Scienze Matematiche, Fisiche e Informatiche, Universit\'a di Parma, Parco delle Scienze 7A, I-43124 Parma, Italy}

\date{\today}
\maketitle

\section{Bulk magnetic measurements}

\begin{figure} 
\centering
\includegraphics[width = 1\linewidth]{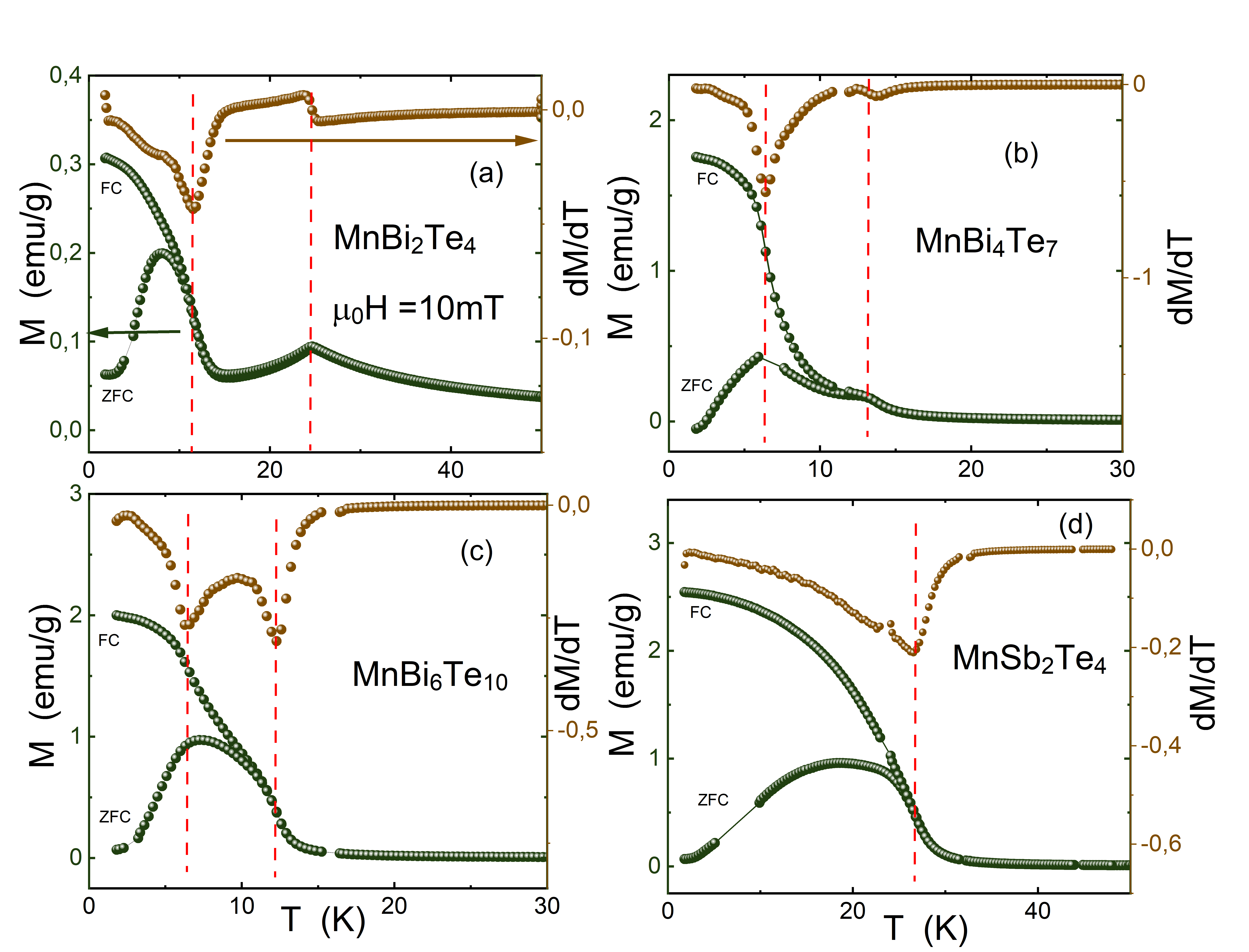}
\caption{Temperature dependence of the magnetization $M(T)$ and of $dM/dT$ in \B $= 10$ mT, for (a) \MBT -$\beta$ , (b) \MBTw, (c) \MBTh~and (d) \MST.}
\label{fig:dcmag}
\end{figure}

A Quantum Design Superconducting Quantum Interference Device  Vibrating Sample Magnetometer (SQUID VSM) was utilized to conduct bulk DC magnetic measurements on the same samples employed for the NMR and \muSR~experiments. The temperature dependence of the magnetization is shown in Fig. \ref{fig:dcmag}, for both zero-field cooled (ZFC) and field-cooled (FC) protocols, as measured over the temperature range 1.8 K to 60 K. 

In the \MBT~sample, for $\mu_0H$= 10 mT, a cusp in $M(T)$ is observed at \TN~$\simeq$ 24.6K (see Fig.~ \ref{fig:dcmag}a, in coincidence with a peak in $dM/dT$), indicating the onset of long-range antiferromagnetic order. Upon further cooling, both ZFC and FC curves exhibit an anomaly, appearing first as an upturn below 15 K, while below 11.5 K the ZFC and FC curves separate, denoting the onset of a ferromagnetic-like character at \Tstar~= 11.5 K, as indicated by a dip in $dM/dT$. This second transition for \MBT\ is suppressed in higher applied fields around 0.5 T, which shows the field-induced character of the transition.

A similar double transition is evidenced by the double dip in $dM/dT$ observed in \MBTw~and \MBTh~as well (see Fig.~\ref{fig:dcmag} b,c), although the AFM cusp is not directly detected in the latter sample. All the transition temperatures are listed in  Tab.~\ref{tab:transition}. Notably, both transition temperatures observed in bulk magnetization are consistent with the corresponding features observed in \muSR, indicating that the anomaly at \Tstar~corresponds to a true thermodynamic phase transition. 

\begin{figure}[!h]
\centering
\includegraphics[width = 0.7\linewidth]{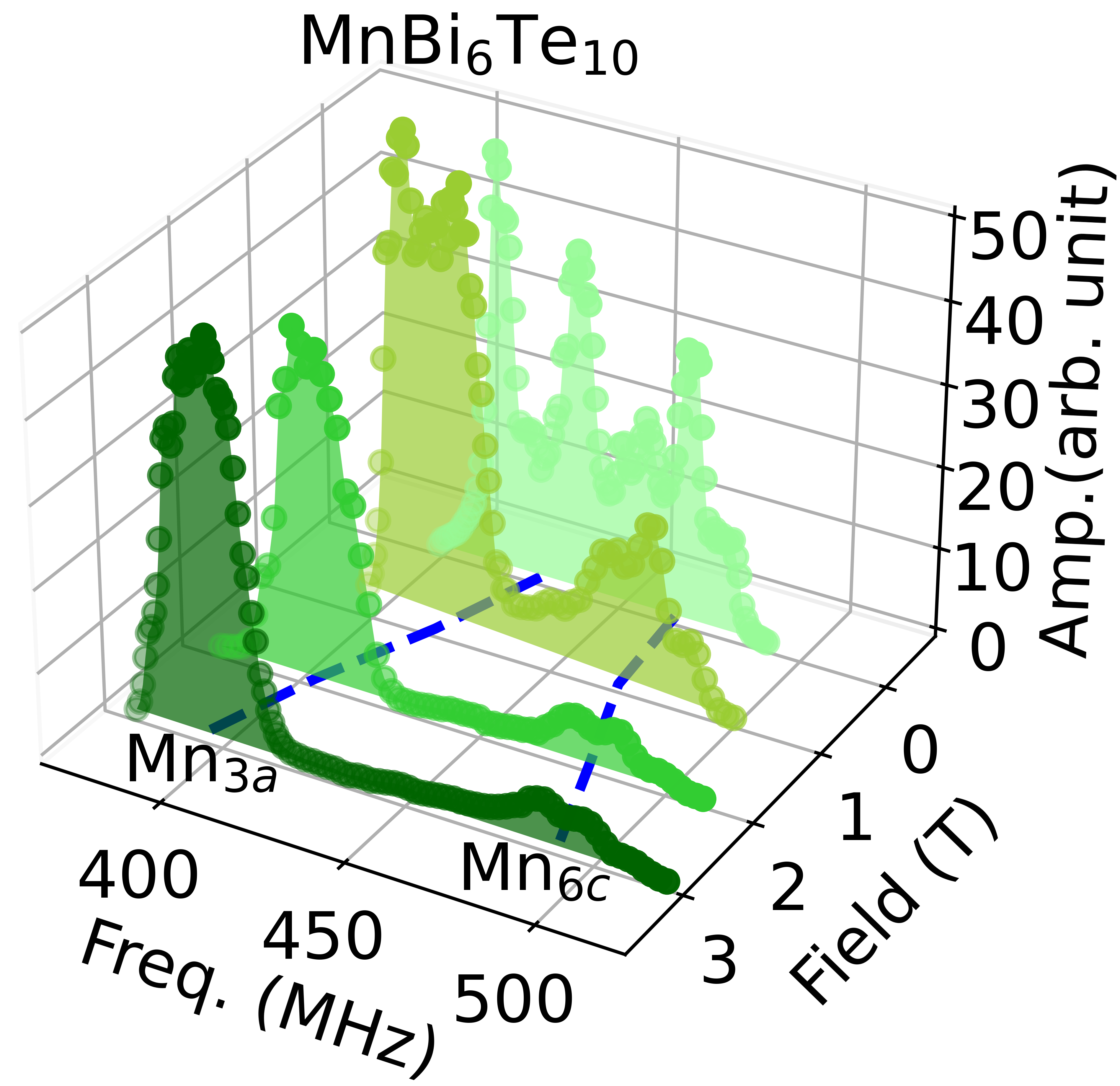}
\caption{\Mn\ NMR spectrum of  \MBTh\ at $T=1.4$ K in increasing applied fields, starting from ZF. Blue dashed lines show the field dependence of the mean frequency for the \Mna\ and \Mnc\ peaks.}
\label{fig:NMR16}
\end{figure}
In the \MST{} sample, a single ferromagnetic-like transition at $T_C$= 26.6K (see Fig.  \ref{fig:dcmag} d) was identified, and the two-transition behavior observed in the \MBT~samples was absent. This absence is attributed to the dominant ferromagnetic signal stemming from \Mna~in the bulk.

\begin{table}[!ht]
    \centering
    \begin{tabular}{|c|c|c|c|c|}
    \hline
    & \multicolumn{2}{c|}{\Tm~[K]} & \multicolumn{2}{c|}{\Tstar~[K]}\\
    Sample  &  SQUID & \muSR & SQUID & \muSR\\
    \hline
        \MBT~$\beta$ & 24.6(1) & 24(1) & 11.5(1) & 12(1)\\
        \MBTw & 13.7(1) & 13(1)  & 6.4(1) & 6(1)\\
        \MBTh & 12.3(1) & 12(1) & 6.5(1)  & 6(1)\\
        \MST & 26.6(1) & - & - & -\\
          \hline
    \end{tabular}
    
    \caption{Transition temperatures observed from both bulk magnetization and \muSR.}
    \label{tab:transition}
\end{table}
\section{NMR}
\subsection{Field dependence of \Mn~NMR in \MBT}
The dashed lines on the horizontal plane in Fig.~1 a, main paper, show that the mean frequencies of the NMR \Mna~and \Mnc~ peaks shift slightly with applied field. These frequencies are plotted as red stars in Fig.~1 g, h, for  \Mna~and \Mnc~ respectively.  Their slope can be appreciated by eye, an effective $\gamma = d\nu/dH\,/\mu_0 $ that vanishes for $\mu_0H\le 1$ T as predicted by Eq.~1 for the AFM case. Indeed, the hyperfine field in an AFM polycrystalline sample, depicted as red arrows in the cartoon (Fig.~1 f) adds in all possible crystal grain orientations to the external field (blue arrow), leading to different total local fields \Bhf~(green arrows), evenly distributed in modulus about the external field value. Recalling that ${^{55}\nu}={^{55}\gamma}|{^{55}\mathbf{B}}|$ this leads to a vanishing shift in first order. The absolute value of $\gamma$ increases around 2 T, as expected in a canted antiferromagnet (CAFM) polycrystal. Indeed, a crystal grain whose hexagonal $c$ axis forms an angle $\theta$ with the field undergoes a spin flop transition at $2\lesssim \mu_0H(\theta)\le 3.57$ T. \cite{lee2019spin,bac2022topological,sass2020magnetic} The powder average results in a distribution of ${^{55}B}$~values (like in the pure AFM case), albeit with a mean shift, due to the coupling to the field ($\gamma<0$ for $\uparrow$\Mna, $\gamma>0$ for $\downarrow$\Mnc).

For comparison, in the FIM case the spin alignment to the field is complete at any orientation and the corresponding \Mna~(Fig.~1 g) and \Mnc~(Fig.~1  h) effective $\gamma$ values  for \MBTw~and \MBTh~(black and blue symbols, respectively) coincide with $\pm{^{55}\gamma}$ within errors, as expected for the $\uparrow$ and $\downarrow$ sublattice respectively.

\subsection{\Mn~NMR spectra of \MBTh}

A 3D view of the spectra for our polycrystalline sample of \MBTh~in the frequency range 350 to 500 MHz, is shown in Fig.~\ref{fig:NMR16}, with ZF at the back and increasing fields \B\ towards the front.

\section{\label{sec:NMR-methods}  DFT simulation of  \Mntitle-NMR spectra}

\begin{figure}
\centering
\includegraphics[width = 1.0\linewidth]{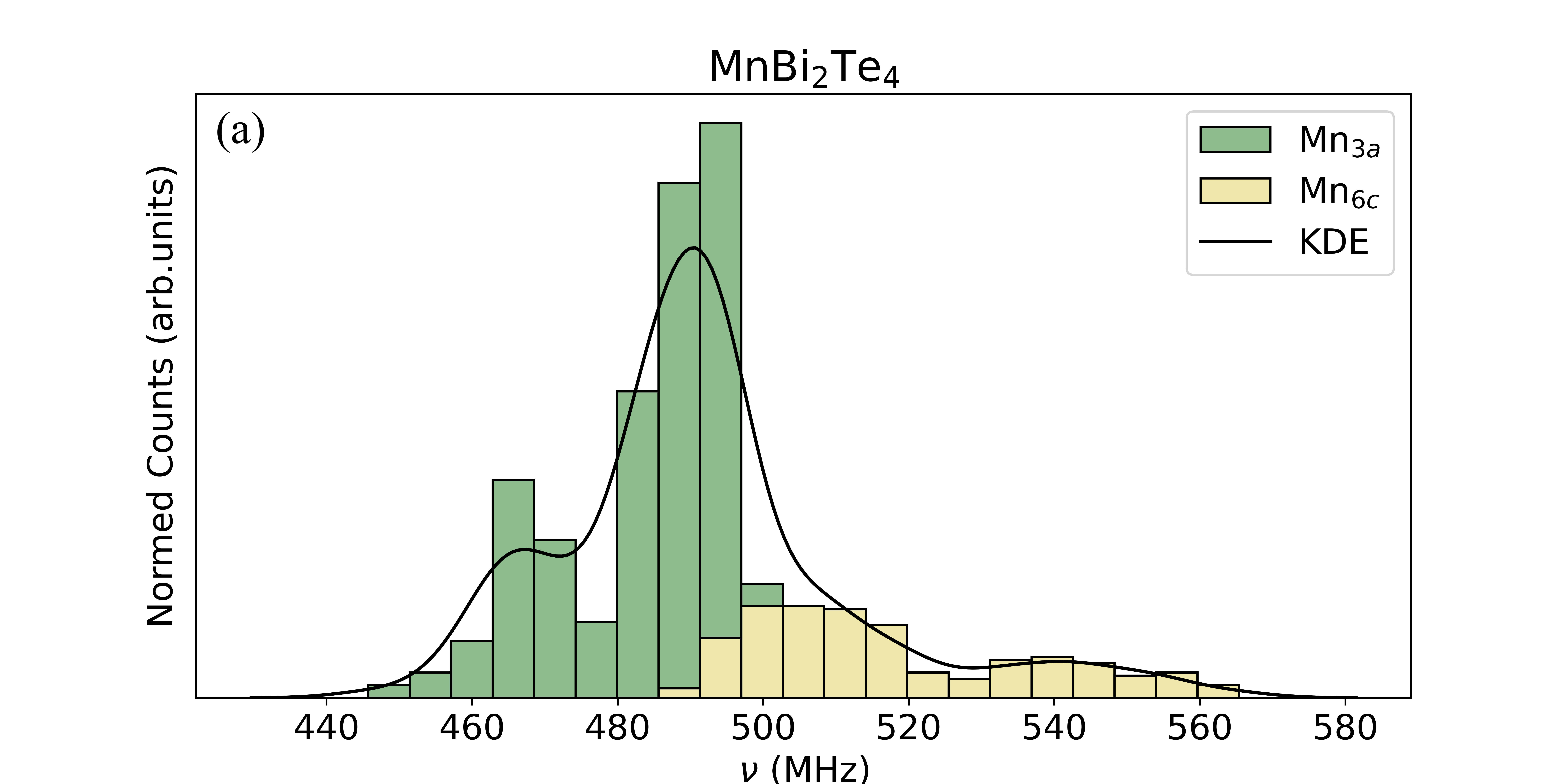}
\includegraphics[width = 1.0\linewidth]{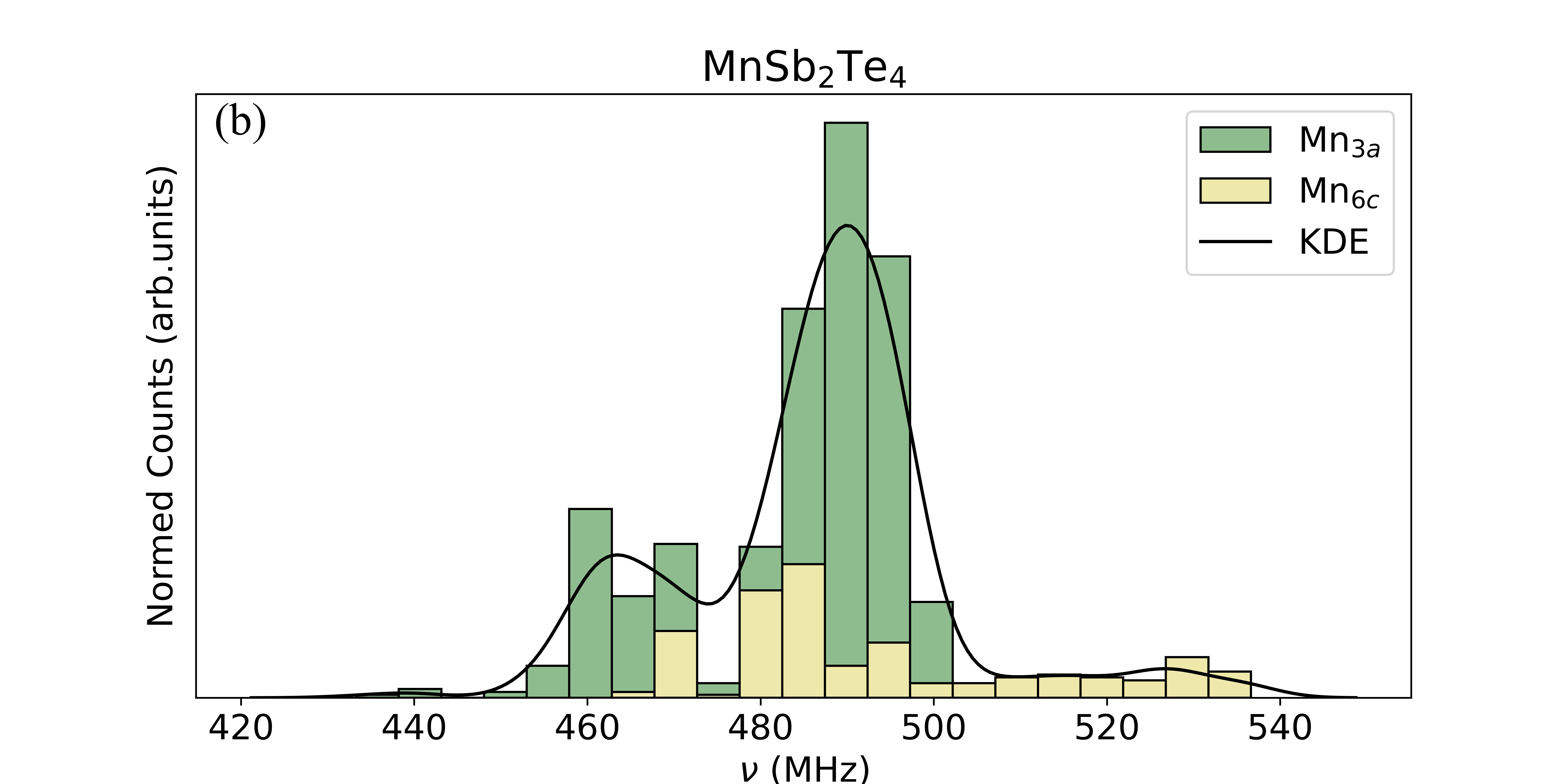}
\caption
{\Mn-hyperfine spectra calculated by DFT in \MBT~ (a) and \MST\ (b), distinguishing the contributions from \Mna~(green bars) and \Mnc~(yellow bars). In the calculation, the \Mnc~moment is assumed antiparallel to the moment of the nearest neighbor \Mna\ in both cases.
 The solid black line is the probability density distribution estimated by the Kernel density function (KDE). }
\label{fig:mnnmr}
\end{figure}

We have calculated 
the hyperfine field at the Mn nucleus for both \MBT~and \MST, first, without site intermixing, by an all electron DFT+U protocol as implemented in the ELK code \cite{elk}.  
We have used a dense 15$\times$15$\times$3 Monkhorst-Pack Mesh for \MBT\ and  15$\times$15$\times$7 for \MST. The  muffin-tin radius of 2.4 a.u. for Mn, 2.8 a.u. for Bi, 2.6 a.u.  for Sb  and 2.6 a.u. for Te were used. The maximum length or cutoff for the $\mathbf{G} + \mathbf{K}$ vectors is 8.0 a.u., divided by  the average of the muffin-tin radii.  The hyperfine field at the  \Mna\ nucleus in the ideal \MBT\  and \MST, calculated within the scalar relativistic approach, is 42.2 T and  45.6 T, corresponding to NMR frequencies of 446 MHz and 482 MHz, respectively. These values are in reasonable agreement with the experiments,  being within 15\% of the values from Fig.~1 g,i (430 and 423 MHz, respectively). For \MST, the fully relativistic approach,  better accounting for spin orbit coupling, yields a magnitude of almost coincident value of 45.5 T, showing that the dominant contribution is the Fermi contact term. 

 We have used the GIPAW code\cite{GIPAW} to model the intersite mixing in \MBT~and \MST, and generate the distribution of \Mn\ hyperfine fields at the \Mna~and \Mnc~nuclei. The calculation details for muon calculations in Supplementary Sec.~\ref{mupart} were adopted.  Hundreds of different structural configurations were generated in a 2$\times$2$\times$2 supercell of the magnetic primitive cell, according to typical values of the Mn/Bi occupancies for the Wyckoff positions $6c$ (94\% Bi and 6\%Mn) and $3a$ (74\% Mn, 21\% Bi and 5\% voids) \cite{Zeugner.cm2019}, by making use of a supercell code \cite{supercell}.  Of all the generated configurations, one hundred were randomly selected for the DFT calculations of the hyperfine field, involving a total of 800 distinct Mn sites. 
 \begin{figure}
\centering
\includegraphics[width = 0.9\linewidth]{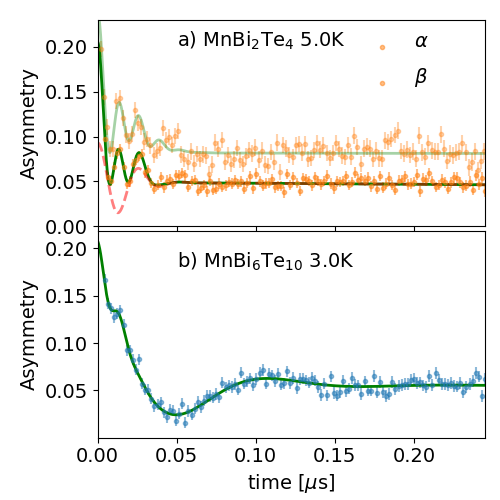}
\caption
{Best fits to Eq. 2, at low temperature for a) \MBT\ $\alpha$ and $\beta$ samples (dashed curve: best fit with components 1,3 set to zero amplitude to highlight component 2), and b) \MBTh, with reduced $\chi^2 = 1.00, 0.98, 0.98$ respectively.}
\label{fig:5K}
\end{figure}
 The resulting distributions of hyperfine fields are shown in Fig.\ref{fig:mnnmr}.

\section{ZF \muSRtitle~fits}
\begin{figure}
\centering
\includegraphics[width = 0.8\linewidth]{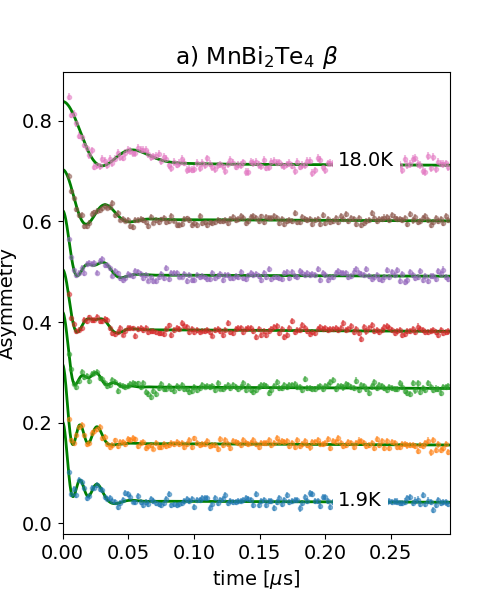}
\caption
{Best fits to Eq. 2, at several temperatures for \MBT~$\beta$ sample.}
\label{fig:multiplot}
\end{figure}

The quality of the fits by Eq.~2 can be better judged on representative sets in Fig.~\ref{fig:5K}. For \MBT~at $T=5$ K the fast initial damping is due to component 1, the visible oscillation is due to component 3, while the contribution of component 2, not so evident in the total best fit, is highlighted by the dashed red curve, where  best fit components 1 and 3 have been set to zero amplitude.  For \MBTh~at $T=3$ K component 1 is missing, component 2 is the large amplitude lower frequency oscillation and component 3 is the lower amplitude fast initial oscillation.  Figure \ref{fig:multiplot} shows that the same good fit quality is obtained at all temperatures by Eq.~2 also for \MBT~$\beta$ (\MBT-$\alpha$ is in Fig.~2 a).

\section{ \label{mupart}  DFT calculation of the muon implantation site and coupling}%

\subsection{Methods}
We determine the muon implantation sites in \MBT~and \MST~by performing collinear spin-polarized density functional theory (DFT) calculations by means of the Quantum Espresso code~\cite{qe2009}. We use the projector augmented pseudopotentials ~\cite{blochl1994} and the semi-local  generalized-gradient-approximation (GGA)  with the  Perdew-Burke-Ernzerhof (PBE) ~\cite{pbe1996} functional and the DFT-D3 van der Waals dispersion energy-correction method \cite{grimme2010} to capture more accurately the effects of the long-range interactions in the structural optimization.  The role of electron-electron interaction on the Mn$-d$ orbitals are considered  within the DFT+U scheme ~\cite{dudarev1998,cococcioni2005,kulik2006,timrov2018}, where we have used a value of $U_{\mathrm{eff}} = (U - J) = 4.0$ eV~\cite{ping2021}.
\begin{figure}[!ht]
\centering
 \includegraphics[width = 0.98\linewidth]{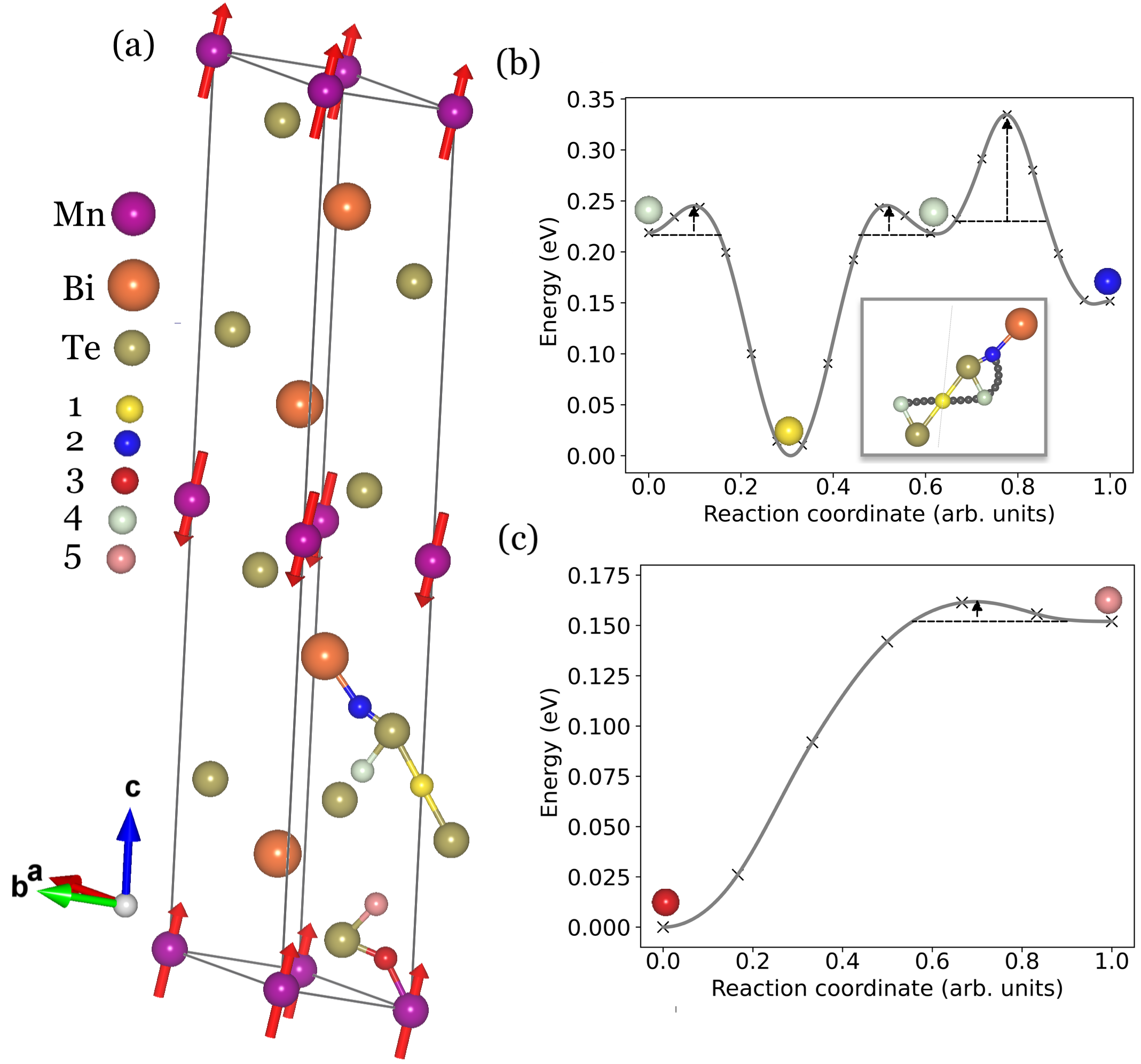}
\caption{(a) Muon sites  1-5 shown in the primitive magnetic cell of \MBT (calculations were performed in a 2$\times$2$\times$1 supercell). Nudge elastic band (NEB) minimum energy landscape along two paths connecting calculated muon sites (b) polyline 4$'$-1-4-2 (site 1 is a high symmetry position and is flanked by sites 4 and its symmetry equivalent, 4$'$, Inset shows the structural path, where black spheres represents sites between the NEB path), and (c) straight 3-5.}
\label{fig:mbtsiteneb}
\end{figure}

The DFT+U relaxations were performed in an interlayer antiferromagnetic cell (primitive cell doubled along c)  for \MBT{}\cite{Otrokov2019}.  
To accommodate the effect of the spurious artificial interactions of the positive muon impurity modeled with a hydrogen pseudopotential, we have used a 2$\times$2$\times$1 supercell, positively charged, with a uniform charge background to restore charge neutrality.   The plane-wave kinetic energy and charge density cut-offs used are 70 Ry and 630 Ry, respectively. For structural optimizations, a 3$\times$3$\times$1 k-point mesh was used for the 57-atom supercell, while it was doubled along all the axes for the calculation of the contact hyperfine field via a self-consistent calculation. The atomic positions were optimized till force and energy thresholds of 1$\times$10$^{-3}$ a.u(Ry/Bohr) and 1$\times$10$^{-4}$ Ry while the lattice parameters remain fixed at the DFT+U obtained values (i.e., without hydrogen).

\begin{table}

    \centering
    \caption{DFT relaxed  fractional coordinates in the primitive magnetic unit cell of \MBT\ both for the atomic positions (in half of the magnetic cell) and the calculated muon sites, labelled 1 to 5. The lattice parameters are as follows: $a= b= 4.3795$ \AA, $c=27.6430$ \AA, $\alpha= \beta= 80.8841^\circ$, $\gamma = 60.0^\circ$.}
    \label{tab:sites}
    \begin{tabular}{| c | c |}
    \hline
         Atom &  Coordinates \\
          \hline
         Mn &  (0.0, 0.0, 0.0)\\
         Bi & (0.4233, 0.4233, 0.3651)\\
         Bi & (0.5767, 0.5767, 0.1349),\\
         Te(1) & (0.1321, 0.1321, 0.3018)\\
         Te(1) & (0.8679, 0.8679, 0.1982)\\
         Te(2) & (0.2946, 0.2946, 0.0582)\\
         Te(2) &(0.7055, 0.7055, 0.4418)\\
         \hline
         $\mu$ site &  Coordinates\\
          \hline
         1 & (0.0000, 0.0001, 0.2499) \\
          2 & (0.2697, 0.2697, 0.3191) \\
         3 &  (0.1134, 0.1134, 0.0499)\\
         4 & (0.2461, 0.2461, 0.2495)\\
         5  & (0.1333, 0.1884, 0.1077)\\
         \hline
    \end{tabular}

\end{table}

\begin{table*}
\begin{threeparttable}
\caption{\MBT: muon bond labels and muon sites 1-5  ($\overline{3,5}$ represents the average of 3, 5), DFT total energy difference ($\Delta E$) to the lowest energy and local fields in mT; \Bc~contact (along the bulk magnetization), \Bd~dipolar, and \Bmu~total fields in mT. All fields are rounded to the mT. Magnetic moments on Mn are aligned along the Cartesian z axis ([0001] direction) \cite{Yan.prm2019} in the primitive cell. Top section: \MBT~muon sites without  Mn-Bi intermixing; bottom section:  same muon sites affected by the intermixing.}
    
    \label{tab:fields}
    \begin{tabular*}{\textwidth}{@{\extracolsep{\fill}} l *{6}{c}| }
    
          \hline
          \hline
         Label & index &  $\Delta{E}$  &  \Bc  & \Bdip   & \Bmu  \\
         &&(meV)&&&\\
          \hline
          &1   & 0            & 0                   &   (0, 0, 0)      & 0     \\
        &4  & 220 & -152   & (0, 0, -1)  & 153 \\
        & $4'$\tnotex{tnxx87}  & 220 & 152   & (0, 0, 1)  & 153 \\
      \TT & $\overline{4,1,4'}$  &  & 0   & (0, 0, 0)    & 0 \\
       \\
        &&&&&\\
       Te-$\mu$-Bi & 2  & 150 &  92   & (2,  1, 0)         & 93 \\
        &&&&&\\
         & 3  & 200 & -497  & (463,  267, -86)            & 791 \\
          & 5  & 350  & 394    & (42,  23, -44)       & 354\\ 
         Te-$\mu$-Mn & $\overline{3,5}$  &   &  -52    & (253,  145, -65)       & 314\\          \hline
         \multicolumn{6}{c}{ } \\
         \multicolumn{6}{l}{ With antisite-mixing} \\
         \multicolumn{6}{c}{ } \\
         \hline
        Label & index &  nn &  \Bc  & \Bdip    & \Bmu  \\
         &&antisite&&&\\
         \hline
        \TBM & 2  & \Mnc &  -282   & (443, 256, 156)         & 527 \\
       &&&&&\\
      
       \TMB & $\overline{3,5}$  &  \Bia  &  236   & (-62,  -44, -293)        &95\\ 
      &&&&&\\

       \TMM\tnotex{tnxx86}~~ & $\overline{3,5}$  & \Mnc  & 288    & (-2,  -64, -8)        &287\\ 
          \hline
            \hline
          
\end{tabular*}
     \begin{tablenotes}
           \item[a]  \label{tnxx87} Index $4'$ is the symmetry equivalent  site of index 4 illustrated in the NEB path, Fig.\ref{fig:mbtsiteneb}b.
            \item[b]  \label{tnxx86}These are the\TM~sites with the closest \Bic~replaced by \Mnc~(see Fig~\ref{fig:mbtsite2}e).
		\end{tablenotes} 
\end{threeparttable}
  \
\end{table*}

 \subsection{\label{sec:MBT}  \MBT~muon sites} 
 We start considering \MBT~without site intermixing. Structural optimization including accurate treatment of the van der Waals interactions produce five low total energy candidate muon positions, indexed 1, 2, 3,  4 and 5 in Table~\ref{tab:sites} and Fig.~\ref{fig:mbtsiteneb}. Muon site 1 has the lowest total DFT energy, taken as the origin of the energy scale.
The reported total DFT energy of the other sites are actually the difference $\Delta E$ with this reference energy. They are listed in the upper section of Table \ref{tab:fields}, together with the local field contributions at each site.

\begin{figure}
\centering
\includegraphics[width = 0.98\linewidth]{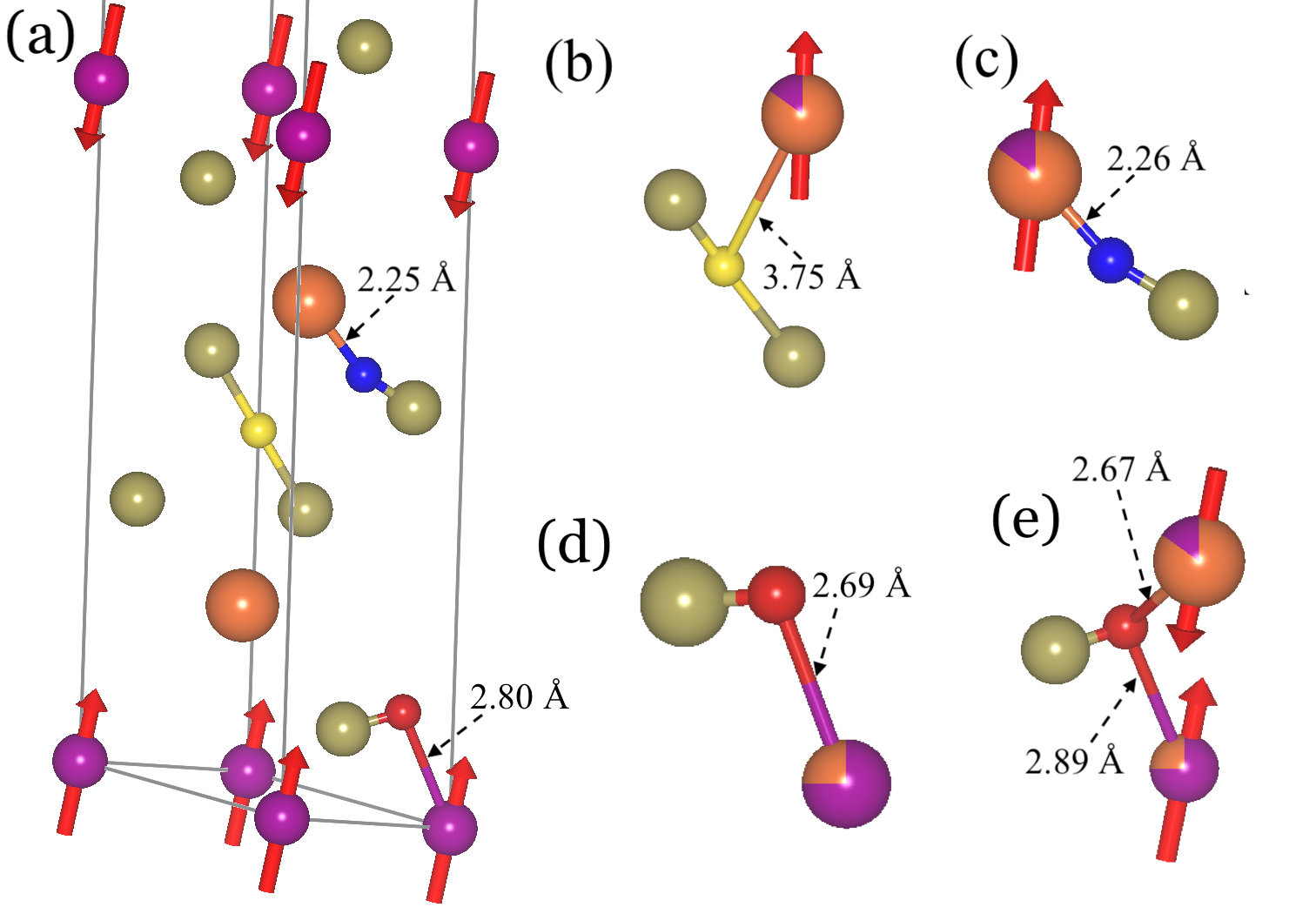}
\caption{Muon sites in \MBT{} (a) labelled Te-$\mu$-Te (yellow sphere), Te-$\mu$-Bi (blue sphere) and Te-$\mu$-Mn (red sphere). The Mn (at 6c) and Bi (at 3a) site mixing has been considered for (b) the next nearest 6c site of Te-$\mu$-Te, (c) the 6c site of Te-$\mu$-Bi (labelled \TBM), (d) the 3a site of  Te-$\mu$-Mn  (labelled \TMB), and (e) the next nearest 6c site of  Te-$\mu$-Mn (labelled \TMM). The value of the $\mu$-Mn/Bi distance is also shown.}
\label{fig:mbtsite2}
\end{figure}
All sites have energies within 0.35 eV, suggesting that all of them are likely to be occupied, although a low energy barrier between close enough sites, or a flat potential landscape may lead to partial delocalization, i.e.~the muon wave function could be broad, spreading over the flat potential region. To investigate the stability of these candidate positions, we have performed nudged elastic band (NEB)~\cite{neb,cineb} calculation, mapping the energy landscape along straight muon paths or polylines across close sites (typically, less than 2 \AA~apart).

The results are shown in Fig.~\ref{fig:mbtsiteneb} c for straight path 3-5. Considering that the zero point energy of muons is typically \EZ~$\approx$ 0.5 eV, the much smaller barriers in the energy landscape shows that muons must partially delocalize over both sites 3 and 5, experiencing an average interaction.  Similar occurrence happens for the minimum energy path (MEP) of 4$'$-1-4-2, \ref{fig:mbtsiteneb} b, where from the energy landscape, two distinct muon sites are observed, one among the 4$'$-1-4 region and the other at site 2. The small energy barrier between 4 and 1 and between the symmetric equivalent 1 and 4$'$, indicates partial delocalization of the muon over 4$'$-1-4. 

In the following analysis we consider only three main muon sites, labelled  \TT~(delocalized over sites 1,  4 and 4$'$),  \TB~(site index 2)  and \TM~(delocalized over sites 3 and 5) according to their chemical bonds. However, the \MBT{} sample is characterized by the presence of \Bia~and \Mnc~antisites. This generates a large number of local configurations. Guided by experiments we approximate it distinguishing three components with widths smaller than the separation of their means, determined in first approximation by the presence or absence of a nn antisite. 

Therefore, we additionally consider the effect of nearest neighbor (nn) antisites on \TB~and \TM~muon sites. They are listed in the second section of Tab.~\ref{tab:fields} as \TBM~(where we consider the \TB~site and \Bic~is replaced by \Mnc), \TMB~(where we consider the \TM~site and \Mna~is replaced by \Bia) and \TMM~(where we consider the \TM~site and the closest \Bic~is replaced by \Mnc).
The next 3a or 6c site in \TT~ has a distance above 3.5 Angstrom and the small effect on its local field is neglected, in first approximation. 
These structures and resulting bond distances are show in Figs.~\ref{fig:mbtsite2} b-e.

\subsection{Local fields in \MBT}

In a ZF $\mu$SR measurement, the total local magnetic field at the muon site consists of the following contributions:  $\mathbf{B}_\mu = \mathbf{B}_{\mathrm{C}} + \mathbf{B}_{\mathrm{dip}} + \mathbf{B}_{\mathrm{L}}$,~\cite{Blundell} where  $\mathbf{B}_{\mathrm{c}}$ is the isotropic contact contribution 
originating from the Fermi interaction, which requires a quantum treatment of the electronic wavefunction and has been obtained here with DFT calculations~\cite{onuorah2018}. The last two terms are obtained by the summation of the long-range dipoles in real space using the Lorentz sphere approach, where $\mathbf{B}_{\mathrm{dip}}$ is the dipolar field obtained from the contributions of the Mn magnetic moments within the sphere, while those outside the sphere contribute to the Lorentz term $\mathbf{B}_{\mathrm{L}}$~\cite{bonfa2018}, assuming that the outer sample surface is always unmagnetized in ZF.  For the antiferromagnetic order of \MBT, the $\mathbf{B}_{\mathrm{L}}$  term vanishes. Table \ref{tab:fields} reports our results, where the Mn magnetic moment has been fixed at 4.3 $\mu_B$, the value obtained from our NMR measurements. The non-negligible effect of the muon induced displacements on the host atomic positions is included in calculating \Bd. Due to the approximations inherent in \DFTm, absolute agreement with experimental values is not expected to be better than within 25\%.

Since the energy minima of sites 3 and 5 are rather close, compared to \EZ, a rough estimate of their local field is just the value, \Bmu, obtained from the average of their local components, reported in Tab.~\ref{tab:fields} as $\overline{3,5}$ . This value is actually in close agreement with the experimental field $B_2$ at zero temperature.

For more than one site, namely, \TT, \TB~and \TMB, the predicted local field value is close to zero (within the second moment of the field for component 1, $\Delta B_1$).

Most notable is \TBM, where, due to the short $\mu$-\Mnc~distance $\approx$ 2.3 \AA, (Fig.~\ref{fig:mbtsite2}c) the internal muon field increases considerably to 527 mT (see Table~\ref{tab:fields}). This value agrees very closely with the experimental value of $B_3$ at zero temperature. The site assignment for \MBT~is summarized in Tab.~I.

\subsection{ Local fields in \MBTnntitle, $\mathbf{n=1,2}$}

\begin{figure}
\centering
\includegraphics[width = 0.9\linewidth]{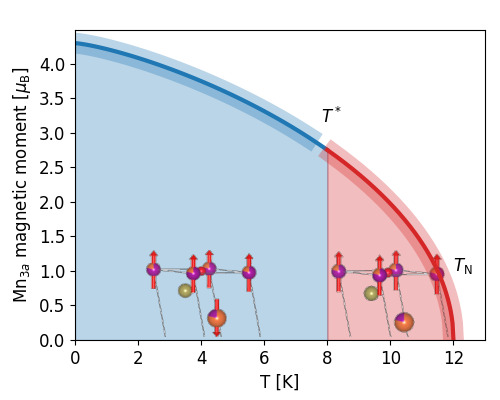}
\caption{\MBTh~temperature dependence of the magnetic moment on \Mna, low temperature value from NMR, temperature dependence from \muSR, $\Delta B_2(T)$. The inset shows the smallest unit cell portion sufficient to highlight the change.}
\label{fig:phasediagram_n}
\end{figure}

 The \MBTnn~structure of the two compounds, is formed by quintuple \BiTe~layers intercalated between the septuple layers of \MBT, therefore we consider the same sites of Tab.~\ref{tab:fields}, plus further muon sites, similar to \TB~and \TM, in the additional quintuple layers. In the FM structure the hyperfine field is never vanishing by symmetry at any site, as it is the case for \TT~in the AFM structure. Still, most of these site are far away from the \Mna~layer and have small local fields. Therefore they contribute to a broad field distribution centered at the Lorentz field value, \BL $\approx 40$ mT ($30$ mT) for \MBTw~(\MBTh). This corresponds closely to the experimental $B_2$ value at zero temperature. Its  temperature dependence is shown in Fig.~\ref{fig:phasediagram_n} for $n=2$, rescaled to the magnetic moment per site with the ZF NMR calibration at 1.4 K, and \Tstar~ is the temperature where the $B_3$ term disappears (see Fig.~2 f). 

 The field value corresponding to site \TM, the highest energy site in \MBT, is not observed experimentally. This is not surprising, in  view both of its higher energy and of the large number of sites contributing to $B_2$. 

\subsection{Local fields in \MST}

\begin{figure}
\centering
\includegraphics[width = 0.7\linewidth]{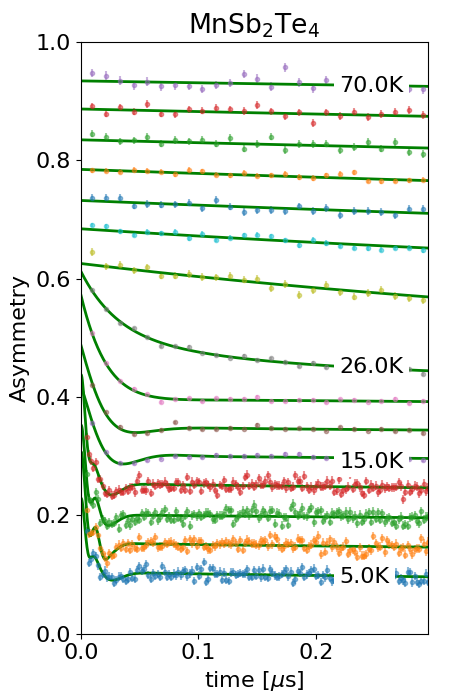}
\caption{\MST~\muSR~asymmetry with best fit to Eq.~2 (see text).}
\label{fig:MSTmulti}
\end{figure}

\begin{figure}
\centering
\includegraphics[width = 0.7\linewidth]{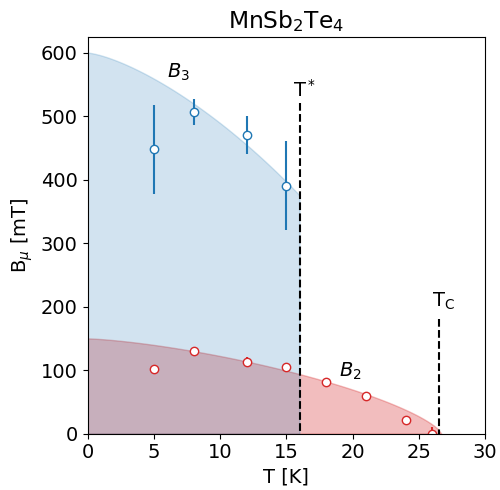}
\caption{Best fit fields $B_2$ and $B_3$ in \MST.}
\label{fig:MSTfields}
\end{figure}
 
Figure \ref{fig:MSTmulti} shows the best fits to Eq.~2, where the overdamped $B_1$=0  transverse component is missing, and Fig.~\ref{fig:MSTfields} shows the best-fit field values vs.~temperature. 

The calculation yields similar results and the muon site classification is similar to that of \MBT, but the actual sample interlayer order is ferromagnetic,

giving rise to a Lorentz contribution \BL $\approx 70$ mT. Here, the difference between the Sb and the Mn radii is even smaller than in the Bi case, therefore the occupancy of the \Mnc~antisite is larger, and all local field distributions become much broader. The $B_2$ component agrees with the prediction for all the sites that have low local field values in \MBT, namely, \TT~, \TB~and \TMB.

\bibliography{references}